\documentclass[final]{siamltex}

\usepackage{euscript,amsmath,amssymb,amsfonts,graphicx,color,cite,hyperref,chngcntr}
\allowdisplaybreaks[1]

\numberwithin{equation}{section}
\counterwithout{figure}{section}

\usepackage{epstopdf,fancybox,subfigure}

\newcommand{\bqq}{\begin{equation}}
\newcommand{\eqq}{\end{equation}}
\newcommand{\bqs}{\begin{equation*}}
\newcommand{\eqs}{\end{equation*}}
\newcommand{\ignore}[1]{}
\newcommand{\e}{{\rm e}}
\renewcommand{\d}{{\rm d}}
\newcommand{\pd}{\partial}

\newcommand{\R}{{\mathbb R}}
\newcommand{\C}{{\mathbb C}}

\newcommand{\D}{\displaystyle}
\newcommand{\mc}{\mathcal }
\newcommand{\ve}{\varepsilon}

\newcommand{\Lo}{{\mathcal L}}

\newcommand{\cH}{\mathcal{H}}

\newcommand{\vv}{{\mathbf{v}}}
\newcommand{\X}{\mathcal{X}}
\newcommand{\calr}{\mathcal{R}}
\newcommand{\F}{\mathcal{F}}
\newcommand{\caln}{\mathcal{N}}

\newcommand{\bp}{\mathbf{\Phi}}

\title{Pulse bifurcations in stochastic neural fields}
\author{Zachary P. Kilpatrick\thanks{University of Houston,
Department of Mathematics,
671 Phillip G. Hoffman,
Houston, TX 77204 US (\href{mailto:zpkilpat@math.uh.edu}{zpkilpat@math.uh.edu}). This author was supported by an NSF grant (DMS-1311755)} \and Gr\'egory Faye\thanks{University of Minnesota,
School of Mathematics,
206 Church Street S.E.,
Minneapolis, MN 55455 US  (\href{mailto:gfaye@umn.edu}{gfaye@umn.edu}). This author was supported by an NSF grant (DMS-1311414).}}

\begin{document}

\maketitle

\begin{abstract}
We study the effects of additive noise on traveling pulse solutions in spatially extended neural fields with linear adaptation. Neural fields are evolution equations with an integral term characterizing synaptic interactions between neurons at different spatial locations of the network. We introduce an auxiliary variable to model the effects of local negative feedback and consider random fluctuations by modeling the system as a set of spatially extended Langevin equations whose noise term is a $Q$-Wiener process. Due to the translation invariance of the network, neural fields can support a continuum of spatially localized bump solutions that can be destabilized by increasing the strength of the adaptation, giving rise to traveling pulse solutions. Near this criticality, we derive a stochastic amplitude equation describing the dynamics of these bifurcating pulses when the noise and the deterministic instability are of comparable magnitude. Away from this bifurcation, we investigate the effects of additive noise on the propagation of traveling pulses and demonstrate that noise induces wandering of traveling pulses. Our results are complemented with numerical simulations.
\end{abstract}



\pagestyle{myheadings}
\thispagestyle{plain}


\section{Introduction} \label{model}
Spatially structured cortical activity serves a variety of functions of the brain \cite{wang10}. For instance, persistent and localized neural activity has been observed in monkey prefrontal cortex during spatial working memory experiments, and the location of this ``bump" represents a remembered location \cite{colby95,compte00}. Bumps are also known to encode an animal's position during spatial navigation tasks \cite{mcnaughton06,yoon13}. In addition, propagating waves of neural activity have been implicated in both motor \cite{rubino06} and sensory \cite{petersen03,ferezou07,xu07} tasks. Typical mathematical models of such large-scale spatiotemporal activity take a mean field approach \cite{coombes05,bressloff12}, rather than modeling biophysical details. Such neural field equations support a rich variety of spatially structured solutions including waves \cite{ermentrout93,pinto01}, bumps \cite{amari77,laing02}, and Turing patterns \cite{bressloff01b}. In their simplest form, these systems are single (scalar) equations describing spatiotemporally coarse grained neural activity whose dynamics are largely determined by their prescribed synaptic connectivity \cite{wilson73}.

In recent years, there has been considerable interest in how the dynamics of neural fields are modified when local negative feedback is incorporated into models \cite{laing01b,pinto01,coombes05b,troy07,kilpatrick10}. Scalar neural fields with symmetric connectivity tend to only support traveling fronts \cite{ermentrout93,bressloff01}, in the case of excitatory connectivity, or stationary bumps and patterns \cite{amari77,laing02}, in the case of lateral inhibitory connectivity. However, once negative feedback is considered, a variety of spatiotemporal dynamics can be found, such as traveling pulses \cite{pinto01}, breathers \cite{folias04}, and spiral waves \cite{huang04}. One key observation concerning traveling pulses is that they arise through one of two typical bifurcations: (a) a ``back" arises on a front \cite{pinto01} or (b) a stationary bump begins to drift \cite{hansel98}. In related situations, perturbative techniques have been used to derive amplitude equations for the speed of traveling fronts \cite{bressloff03} or periodic patterns emerging from the homogenous state \cite{curtu04}. Typically, propagation of pulses occurs when the strength and time-scale of the local negative feedback process are sufficiently large \cite{pinto01,kilpatrick10}, and we plan to take a closer look at this in the present work.

Furthermore, much recent work has been concerned with the effects of noise on spatiotemporal dynamics of neural fields \cite{brackley07,hutt08,bressloff12b}. Techniques have been adapted from the analysis of front propagation in stochastic partial differential equations \cite{schimansky83,armero98,sagues07}. Typically, studies assume noise is weak and use perturbation theory along with solvability conditions to derive effective equations for the stochastic motion of a front \cite{brackley07,bressloff12b} or bump \cite{kilpatrick13,kilpatrick13b} in a neural field. However, Hutt et al. specifically consider the effect of noise close to a Turing instability, using a stochastic center manifold calculation to show noise shifts the bifurcation point \cite{hutt08}. We plan to build on these studies by considering a stochastic neural field with local negative feedback, capable of supporting traveling pulse solutions. In particular, we will explore the effects of noise close to and away from a pitchfork bifurcation that represents the transition from stationary bumps to traveling pulses.

Local negative feedback is considered by introducing an auxiliary variable that represents spike frequency adaptation \cite{pinto01,folias04,coombes05b} or synaptic depression \cite{york09,kilpatrick10}. One model of such negative feedback assumes the auxiliary variable depends linearly on synaptic input \cite{pinto01,laing01b}
\begin{subequations} \label{ringad}
\begin{align}
\frac{\pd u(x,t)}{\pd t} &= - u(x,t) - \beta v(x,t) + \int_{- \pi}^{\pi} w(x - y) f(u(y,t)) \d y, \label{ringada} \\
\frac{\pd v(x,t)}{\pd t} &=  \alpha [u(x,t) - v(x,t)].  \label{ringadb}
\end{align}
\end{subequations}
The synaptic drive $u(x,t)$ to the neural population at position $x$ is augmented by subtractive negative feedback $v(x,t)$ with strength $\beta$, evolving at rate $\alpha$. The strength of synaptic connections is described by the kernel $w(x-y)$, typically considered to be an even symmetric function \cite{hansel98,ermentrout98,bressloff02,bressloff12}. To demonstrate simply in examples, we will employ the cosine weight function
\begin{align}
w(x-y) = \cos (x-y),  \label{cos}
\end{align}
accepted as an approximation to the lateral inhibitory connectivity in sensory cortical networks \cite{benyishai95,somers95}. However, our results do apply widely to general $w$. The relationship between synaptic drive $u$ and output firing rate is described by the nonlinearity $f(u)$. Typically in neural fields, this is taken to be a sigmoidal function \cite{coombes05,bressloff12}
\begin{align}
f(u) = \frac{1}{1 + \e^{- \gamma (u- \theta)}},  \label{sig}
\end{align}
with gain $\gamma$ and threshold $\theta$, and often analysis is eased by considering the high gain limit $\gamma \to \infty$ so \cite{amari77}
\begin{align}
H(u- \theta) = \left\{ \begin{array}{cl} 1 & : u\geq\theta, \\ 0 & : u< \theta, \end{array} \right. \label{H}
\end{align}
the Heaviside function. We will show our results apply to the general sigmoid (\ref{sig}), but will demonstrate them for the Heaviside (\ref{H}) as well.

It is useful to note that (\ref{ringad}) can be expressed as a single second-order integro-differential equation by first differentiating (\ref{ringada}) so
\begin{align*}
\frac{\pd^2 u(x,t)}{\pd t^2} &= - \frac{\pd u(x,t)}{\pd t} - \beta \frac{\pd v(x,t)}{\pd t} + \int_{- \pi}^{\pi} w(x - y) f'(u(y,t)) \frac{\pd u (y,t)}{\pd t} \d y,
\end{align*}
and then substituting this and (\ref{ringada}) into (\ref{ringadb}) to yield
\begin{align}
{\mc Q}u(x,t) &= \int_{- \pi}^{\pi} w(x-y) \left[ f'(u(y,t)) \frac{\pd u(y,t)}{\pd t} + \alpha f(u(y,t)) \right] \d y,  \label{secordeq}
\end{align}
where ${\mc Q}$ is the second-order linear operator
\begin{align}
{\mc Q}u(x,t) = \frac{\pd^2 u (x,t)}{\pd t^2} + (1 + \alpha) \frac{\pd u(x,t)}{\pd t} + \alpha ( 1+ \beta) u(x,t).  \label{sectempop}
\end{align}
Here, we will study how the pitchfork bifurcation and propagation of traveling pulses in (\ref{ringad}) and equivalently (\ref{secordeq}) is modified by noise.

To do so, we will consider a stochastic neural field equation that describes the effects of adding noise to (\ref{ringad}). Effects of fluctuations will appear in the evolution equation for the adaptation variable $v$ (\ref{ringadb}), resulting in the system of Langevin equations \cite{hutt08,faugeras09b,bressloff12}
\begin{subequations} \label{radlang}
\begin{align}
\d u (x,t) &= \left[ - u (x,t) - \beta v (x,t) + \int_{- \pi}^{\pi} w(x-y) f(u(y,t)) \d y \right] \d t, \label{radlanga} \\
\d v (x,t) &= \alpha \left[ u (x,t) - v (x,t) \right] \d t + \ve \d W (x,t), \label{radlangb}
\end{align}
\end{subequations}
where $\d W (x,t)$ is the increment of a noise process. Throughout this paper, we will assume that $W$ is a $Q$-Wiener process on the Hilbert space of $2\pi$-periodic functions $\cH:=L^2_{per}([0,2\pi],\R)$ where $Q:\cH\rightarrow \cH$ is a non-negative, symmetric bounded operator on $\cH$ such that $\text{Tr}(Q)<\infty$. From this definition, there exists a complete orthonormal basis $\left( e_k \right)_{k\geq 1}$ and a sequence of positive real numbers $\left(\gamma_k \right)_{k \geq 1}$ such that
\begin{align*}
Q(e_k) =\gamma_k e_k, \hspace{1cm} \sum_{k=1}^{\infty} \gamma_k<\infty, \hspace{1cm} W(x,t)=\sum_{k=1}^{\infty} \sqrt{\gamma_k}B_k(t)e_k(x).
\end{align*}
The family $\left(B_k\right)_{k\geq1}$ consists of independent real-valued standard Brownian motions. It is then a consequence \cite{da08}, that $\langle \d W \rangle = 0$ and
\begin{align}
\langle \d W(x,t) \d W(y,s) \rangle = \mathbf{C}(|x - y|) \delta (t - s) \d t \d s,
\end{align}
where the function $\mathbf{C}$ is related to the bounded operator $Q$ via the representation
\bqs
Q:\left\{
\begin{array}{ccc}
\cH & \rightarrow & \cH, \\
\varphi &\mapsto & Q(\varphi)=\mathbf{C}\ast \varphi,
\end{array}
\right.
\hspace{1cm} \mathbf{C}\ast \varphi(x)=\int_0^{2\pi}\mathbf{C}(|x-y|)\varphi(y)dy.
\eqs
The existence and uniqueness of mild solutions to \eqref{radlang}, for a given initial condition $(u_0,v_0)\in \cH \times \cH$, with trace class noise $W$ defined as above, is guaranteed under the Lipschitz condition on $f$ and the fact that the operator $K$ defined as
\bqs
K:\left\{
\begin{array}{ccc}
\cH & \rightarrow & \cH, \\
\varphi &\mapsto & K(\varphi)= w\ast \varphi,
\end{array}
\right.
\eqs
is compact on $\cH$. If we denote $\mathbf{A}$ the following matrix
\bqq
\label{MatrixLin}
\mathbf{A}=\left(\begin{matrix} -1 & -\beta \\ \alpha & -\alpha \end{matrix} \right),
\eqq
then mild solutions of \eqref{radlang} satisfy the equation
\bqq
\label{mild}
\left(\begin{matrix} u  \\ v  \end{matrix} \right)=e^{\mathbf{A} t}\left(\begin{matrix} u_0  \\ v _0 \end{matrix} \right)+\int_0^te^{\mathbf{A}(t-s)}\left(\begin{matrix} w\ast f(u(s)) \\ 0 \end{matrix} \right)ds+\int_0^t e^{\mathbf{A}(t-s)}\left(\begin{matrix} 0 \\ \epsilon \d W(s) \end{matrix} \right).
\eqq
The proof of existence and uniqueness can be found in  \cite{riedlerkuhen13,faugeras13} which relies on the theory of stochastic differential equations for $Q$-Wiener processes developed in \cite{da08}. By solutions of \eqref{radlang} we will always mean mild solutions of \eqref{mild}.

It is also important to explain why we only consider noise in the adaptation variable. First, cortical adaption typically occurs on a much slower timescale than the synaptic dynamics described by the activity variable $u(x,t)$ \cite{tsodyks98,benda03}. Second, since one can always recast the pair of equations (\ref{radlang}) as a single second-order evolution equation, as shown for the deterministic case (\ref{secordeq}), noise contributions to both (\ref{radlanga}) and (\ref{radlangb}) can be combined into a single stochastic process. Therefore, the full range of stochastic effects of additive noise in either $u(x,t)$ or $v(x,t)$ can be captured with $\d W(x,t)$. Collecting stochastic effects into a single variable will help make our analysis more transparent. This specific formulation will be useful for studying diffusive behavior of traveling pulses away from bifurcations and deriving amplitude equations near bifurcations.

The paper is organized as follows. First, in section \ref{drift} we show that adaption can generate traveling pulse solutions in a model that also supports stationary localized bump solutions. This is due to the occurrence of a symmetry breaking bifurcation of bump solutions analogous to that found in the case of traveling front solutions in neural fields with adaptation \cite{bressloff03}. More precisely, we show that stationary bumps undergo a pitchfork bifurcation at a critical rate of negative feedback leading to a pair of counter-propagating traveling pulses.  In the following section \ref{centerman}, we use a center manifold approach to study this drift bifurcation and compute leading order expansions for the wave speed and relative position of the bifurcating traveling pulses. We then address in section \ref{stochcm} how these amplitude equations are perturbed when we introduce additive noise into the neural fields. Near this criticality, we can derive a stochastic amplitude equation describing the dynamics of these bifurcating pulses when the stochastic forcing and the deterministic instability are of comparable magnitude. Then, we show in section \ref{existpulse} that well beyond the pitchfork bifurcation traveling pulses persist for the unperturbed system and that they are stable (section \ref{stabpulse}). Finally, in section \ref{wanderpulse}, we examine the effects of additive noise on the propagation of traveling pulses and show noise-induced wandering type of phenomena.

\section{Drift instability of bumps}\label{drift}
We begin by studying how adaptation can generate traveling pulses in a model that also supports stationary bump solutions. In particular, we will study the transition that stationary bump solutions undergo, via a pitchfork bifurcation, that leads to an instability in the eigenmode associated with translations of their position. Although this problem has been analyzed previously \cite{laing01b,coombes12}, it will be helpful to review it here. We will be studying the effect that noise has on dynamics in the vicinity of this bifurcation in our later analysis (section \ref{stochcm}).

To start, we identify stationary bump solutions that exist in the network with adaptation. Assuming a stationary solution $(u(x,t),v(x,t)) = (U(x),V(x))$, we can rewrite the system (\ref{ringad}) as
\begin{subequations}\label{adbump}
\begin{align}
U(x) + \beta V(x) &= \int_{- \pi}^{\pi} w(x - y) f(U(y)) \d y,  \label{adbump1} \\
V(x) & = U(x).  \label{adbump2}
\end{align}
\end{subequations}
We can substitute the expression (\ref{adbump2}) into (\ref{adbump1}) to yield the single equation
\begin{align*}
(1 + \beta) U(x) = \int_{- \pi}^{\pi} w(x - y) f ( U(y)) \d y.
\end{align*}
Since $U(x)$ must be periodic, we can expand it in a Fourier series
\begin{align*}
U(x) = \sum_{k=0}^N A_k \cos (kx) + \sum_{l=1}^{N} B_l \sin (lx),
\end{align*}
where $N$ is the maximal number of eigenmodes needed to fully describe the solution. One can assume there are a finite number of terms in the Fourier expansion for $U(x)$ if the weight function $w(x-y)$ has a finite number of Fourier modes. This is a reasonable assumption because most typical weight functions can be fully described, or at least well approximated, by a few terms in a  Fourier series \cite{veltz10}. Doing this allows us to always construct solvable systems for the coefficients $A_k, B_l$. For an even symmetric weight kernel $w(x-y)$, which depends purely on the difference $x-y$, we can write
\begin{align}
w(x-y) = \sum_{k=0}^N w_k \cos (k(x-y)) = \sum_{k=0}^N w_k \left[ \cos (kx) \cos (ky) + \sin (kx) \sin (ky) \right],   \label{wfexp}
\end{align}
so that
\begin{align}
A_k = \frac{w_k}{1+ \beta} \int_{- \pi}^{\pi} \cos (kx) f(U(x)) \d x, \ \ \  \ \ \ B_l = \frac{w_l}{1+ \beta} \int_{- \pi}^{\pi} \sin ( l x) f(U(x)) \d x.  \label{AkBlsys}
\end{align}
Since the system (\ref{ringad}) is translation and reflection symmetric, so too will be its solutions \cite{amari77}. Thus, we exclusively look for even solutions, so $B_l = 0$ for all $l$, so $U(x) = \sum_{k=0}^N A_k \cos (kx)$, meaning (\ref{AkBlsys}) becomes
\begin{align*}
A_k = \frac{w_k}{1+ \beta} \int_{- \pi}^{\pi} \cos ( k x) f \left( \sum_{k=0}^N A_k \cos (kx) \right) \d x.
\end{align*}
One can use numerical methods to solve for the coefficients $A_k$, but particular functions allow for direct calculation of solutions. In the case that the weight kernel is a cosine (\ref{cos}), we can exploit the identity
\begin{align}
\cos (x - y) = \cos x \cos y + \sin x \sin y \label{cosid}
\end{align}
and require symmetry to find $U(x) = A \cos x$ where the amplitude $A$ is now defined by the nonlinear scalar equation
\begin{align}
A = \frac{1}{1 + \beta} \int_{- \pi}^{\pi} \cos y f( A \cos y ) \d y.  \label{bamp}
\end{align}


Note that $A=0$ is always a solution of \eqref{bamp}. For a sigmoidal firing rate function (\ref{sig}), it was shown in \cite{veltz10} that a condition for the existence of $A \neq 0$ solution of \eqref{bamp} is given by
\bqq
\label{bumpcond}
1=\frac{f'(0)}{2(1+\beta)},
\eqq
which is simply the linearization of equation \eqref{bamp} at $A=0$.  Suppose that the threshold $\theta$ is fixed and that there exists a gain $\gamma_0$ in (\ref{sig}) such that condition \eqref{bumpcond} is satisfied. Around $\gamma_0$, the equation \eqref{bamp} reduces to the following equation
\bqq
\label{amplitudeP}
0=\frac{\gamma-\gamma_0}{\gamma_0}A+\chi A^3,
\eqq
where the coefficient $\chi$ depends on $\theta$ and $\gamma_0$ \cite{veltz10}. The fact that the previous equation does not have a second order term is a consequence of the even parity of the connectivity function. Equation \eqref{amplitudeP} is the normal form of a pitchfork bifurcation. Note that when $A\neq 0$, then for any $x_0 \in \R$, $\tilde{U}(x) =A\cos(x+x_0)$ is also a stationary solution of \eqref{ringad} because of the translational symmetry of the equations. 

\begin{figure}
\begin{center} \includegraphics[width=13cm]{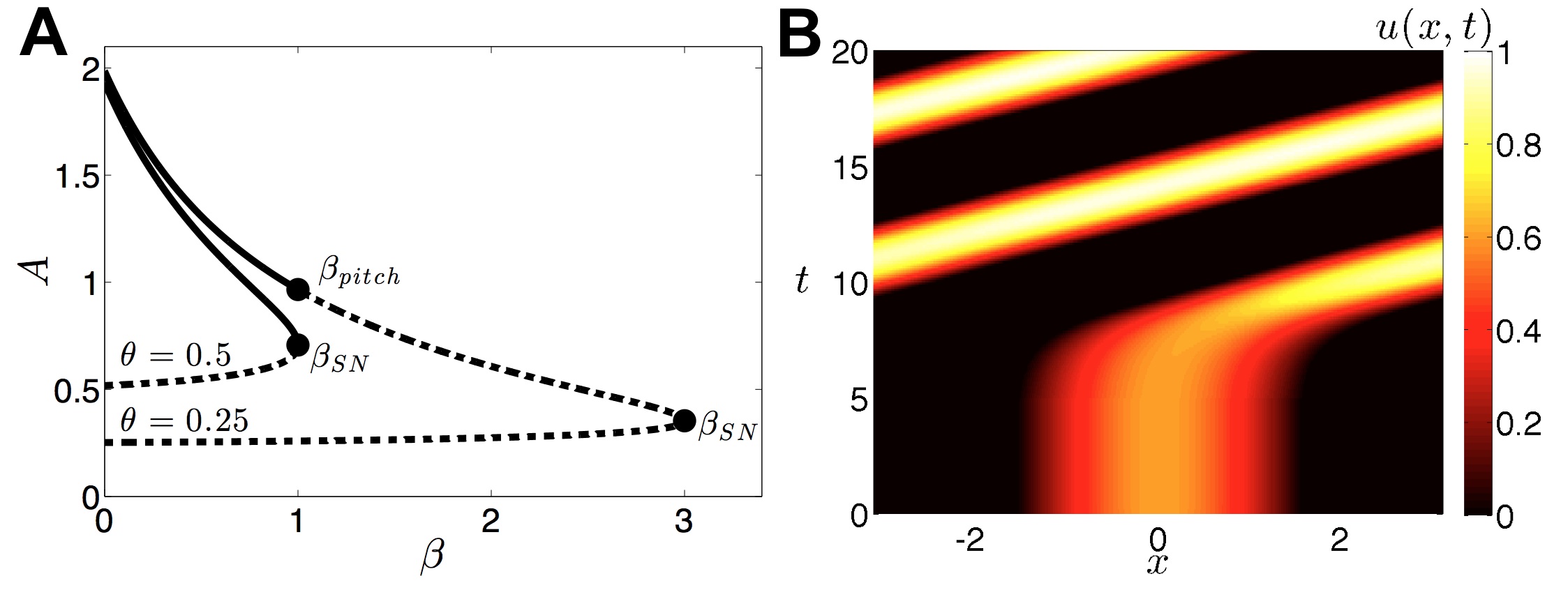} \end{center}
\caption{{\rm ({\bf A})} Plot of the two branches of bump solution amplitudes $A$ as a function of $\beta$. The bottom branch is always unstable (dashed), while the top branch remains stable (solid) until reaching the pitchfork bifurcation at $\beta_{pitch}$. Both annihilate eventually in a saddle-node bifurcation ($\beta_{SN}$). {\rm ({\bf B})} Space-time plot of a bump destabilizing through a drift (pitchfork) bifurcation to form a traveling pulse for $\beta = 2$ and $\theta = 0.25$, perturbed by a rightward shift perturbation at $t=5$. Adaptation rate parameter $\alpha = 1$.}
\label{bwidbif}
\end{figure}

For a Heaviside firing rate function (\ref{H}), we can exactly compute the prefactor $A$ using the equation
\begin{align*}
A = \frac{2}{1 + \beta} \int_0^{\cos^{-1} \theta/ A} \cos x \d x = \frac{2}{1 + \beta} \sin \left( \cos^{-1} \frac{\theta}{A} \right) =  \frac{2 \sqrt{1 - \theta^2/ A^2}}{1+ \beta},
\end{align*}
which we can solve for the bump amplitude
\begin{align}
\label{ampAheaviside}
A = \frac{\sqrt{1 + (1+ \beta) \theta} \pm \sqrt{1 - (1 + \beta ) \theta }}{1 + \beta},
\end{align}
so there are two bump solutions
\begin{align}
U (x) =  \frac{\sqrt{1 + (1+ \beta) \theta} \pm \sqrt{1 - (1 + \beta ) \theta }}{1 + \beta} \cos x,  \label{badsoln}
\end{align}
one of which is always unstable ($-$). Now, the other bump ($+$) will be stable for sufficiently weak adaptation strengths $\beta$. At a critical $\beta$, though, this bump will undergo a drift instability leading to a traveling pulse. The adaptation variable $V$ is then found using the formula (\ref{adbump2}). Note, we can compute the associate half-widths
\begin{align*}
a = \cos^{-1} \left( \frac{\sqrt{1 + (1 + \beta ) \theta} \mp \sqrt{1 - (1 + \beta ) \theta}}{2} \right).
\end{align*}
Note that there is a secondary bifurcation, which is the remnant of the saddle-node bifurcation of the adaptation-free system that occurs when $1 = (1 + \beta ) \theta$. At this point, both bumps vanish. We plot the solution curves for $A$ as a function of $\beta$, along with a typical traveling pulse arising through a drift instability in Fig. \ref{bwidbif}.

We can compute stability by studying the evolution of small, smooth, separable perturbations to the bump solutions. By plugging $(u,v) = (U(x),V(x)) + \ve (\psi (x), \phi (x)) \e^{\lambda t}$ into (\ref{ringad}), Taylor expanding, and truncating to first order, using (\ref{adbump1}) and (\ref{adbump2}) yields the eigenvalue equation
\begin{subequations}\label{adlin}
\begin{align}
\lambda \psi (x) &= - \psi (x) - \beta \phi (x) + \int_{- \pi}^{\pi} w(x - y) f'(U(y)) \psi (y) \d y, \\
\lambda \phi (x) &= \alpha \psi (x) - \alpha \phi (x).
\end{align}
\end{subequations}
We can then expand both spatial functions in Fourier series
\begin{subequations} \label{psiphiexp}
\begin{align}
\psi (x) &= \sum_{k=0}^N {\mc A}_k \cos (kx) + \sum_{k=1}^N {\mc B}_k \sin (kx)  \\
\phi (x) &= \sum_{k=0}^{N} {\mc M}_k \cos (kx) + \sum_{k=1}^{N} {\mc N}_k \sin (kx).  
\end{align}
\end{subequations}
where $N$ is directly determined by the number of terms in the Fourier expansion of $w(x)$. The associated coefficient in (\ref{psiphiexp}) are then determined by the linear system
\begin{align*}
( \lambda + 1) {\mc A}_k + \beta {\mc M}_k &= w_k \int_{- \pi}^{\pi} \cos (kx) f'(U(x)) \psi (x) \d x, \\
(\lambda + 1) {\mc B}_k + \beta {\mc N}_k &= w_k \int_{- \pi}^{\pi} \sin (kx) f'(U(x)) \psi (x) \d x, \\
( \lambda + \alpha ) {\mc M}_k &= \alpha {\mc A}_k \\
( \lambda + \alpha ) {\mc N}_k &= \alpha {\mc B}_k
\end{align*}
which can be reduced to half as many equations by substituting the expressions for ${\mc M}_k$ and ${\mc N}_k$ into the first two equations
\begin{subequations} \label{Qgen}
\begin{align}
\mathbf{Q}(\lambda) {\mc A}_k &= (\lambda + \alpha) w_k \int_{- \pi}^{\pi} \cos (kx) f'(U(x)) \psi (x) \d x, \\
\mathbf{Q}( \lambda ){\mc B}_k &= ( \lambda + \alpha) w_k \int_{- \pi}^{\pi} \sin (kx) f'(U(x)) \psi (x) \d x,  
\end{align}
\end{subequations}
where
\begin{align*}
\mathbf{Q} ( \lambda)  = ( \lambda + \alpha) ( \lambda + 1) + \alpha \beta.
\end{align*}
Solutions of the system (\ref{Qgen}), along with their associated $\lambda$, are the eigensolutions to (\ref{adlin}). We can directly compute the eigenvalues associated with the stability of the bumps in the case of the cosine weight function (\ref{cos}), applying the identity (\ref{cosid}), so
\begin{align*}
\mathbf{Q} ( \lambda ) \left( \begin{array}{c} {\mc A}_1 \\ {\mc B}_1 \end{array} \right) = ( \lambda + \alpha ) \left( \begin{array}{cc} {\mc I} ( \cos^2 x) & {\mc I} ( \cos x \sin x) \\ {\mc I}( \cos x \sin x) & {\mc I} ( \sin^2 x ) \end{array} \right) \left( \begin{array}{c} {\mc A}_1 \\ {\mc B}_1 \end{array} \right),
\end{align*}
where
\begin{align*}
{\mc I} (r(x)) = \int_{- \pi}^{\pi} r(x) f'(U(x)) \d x. 
\end{align*}
First of all, note the essential spectrum is given by $\mathbf{Q}( \lambda^e) = 0$ or
\begin{align*}
\lambda^e_{\pm} = \frac{1}{2}\left[ -(1 + \alpha) \pm \sqrt{(1-\alpha)^2 - 4 \alpha \beta} \right],
\end{align*}
which will surely have negative real part since $\alpha, \beta \geq 0$, so it will not contribute to any instabilities. Now, upon integrating (\ref{bamp}) by parts, we see
\begin{align}
A = \frac{1}{1+ \beta} \int_{- \pi}^{\pi} \cos x f(A \cos x) \d x = \frac{A}{1+ \beta} \int_{- \pi}^{\pi} \sin^2 x f'(A \cos x) \d x. \label{Asin2}
\end{align}
Therefore, as long as $A \neq 0$, the equality (\ref{Asin2}) tells us
\begin{align}
{\mc I} ( \sin^2 x ) = \int_{- \pi}^{\pi} \sin^2 x f'(U(x)) \d x = 1 + \beta.   \label{Isin2}
\end{align}
In the same way, we can derive the identities
\begin{align*}
{\mc I} ( \cos x \sin x ) = 0, \ \ \ \ {\mc I} ( \cos^2 x) = {\mc I} (1) - (1+ \beta ),
\end{align*}
which allow us to show that the eigenvalues indicating the stability of bumps are given by the characteristic equation
\begin{align}
\left| \begin{array}{cc} \mathbf{Q}( \lambda ) - ( \lambda + \alpha) ( {\mc I}(1) - (1 + \beta )) & 0 \\ 0 & \mathbf{Q} (\lambda ) - ( \lambda + \alpha ) ( 1+ \beta )\end{array} \right| &= 0 \nonumber \\
[ \mathbf{Q}( \lambda ) - ( \lambda + \alpha)( {\mc I}(1) - (1 + \beta )) ] [ \mathbf{Q} (\lambda ) - ( \lambda + \alpha )( 1+ \beta ) ] &= 0 \nonumber \\
[\lambda^2 + (2 + \alpha + \beta - {\mc I}(1)) \lambda + 2 ( \alpha + \alpha \beta ) - \alpha {\mc I}(1) ] [ \lambda^2 + ( \alpha - \beta ) \lambda ] &= 0.   \label{badchareq}
\end{align}
Roots of the second quadratic in (\ref{badchareq}) constitute eigenvalues associated with eigenfunctions where $( {\mc A}_1, {\mc B}_1 ) = (0,1)$, odd perturbations to the bump. The first eigenvalue we can identify is $\lambda_0 = 0$, associated with translations of the bump being marginally stable, due to the translation symmetry of (\ref{ringad}). Next, $\lambda_- = \beta - \alpha$ is an eigenvalue that is equivalently zero when $\beta = \alpha$. As we will elaborate upon later, $\lambda_-$ crossing through zero signifies a pitchfork bifurcation, at which the larger bump becomes unstable to translational perturbations. Thus, for strong enough adaptation $\beta$, bumps destabilize to form traveling pulse solutions (Fig. \ref{bwidbif}({\bf B})). Roots of the first quadratic in (\ref{badchareq}) are eigenvalues associated with eigenfunctions where $({\mc A}_1, {\mc B}_1) = (1,0)$, even perturbations of the bump. The wider (narrower) bump is always stable (unstable) to such perturbations. Thus, to study the stability of the wider bump, it suffices to simply study its response to odd perturbations.

\section{Center manifold at the drift bifurcation}  \label{centerman}

In this section, we make precise the formal stability analysis of section \ref{drift} when $\beta=\beta_c:=\alpha$. To this end, we will use a center manifold approach to study this drift bifurcation. Throughout this section, we denote the bifurcation parameter
\bqq
\label{bifpar}
\mu:=\beta-\beta_c=\beta-\alpha.
\eqq 
We first rewrite system \eqref{ringad} as an abstract system on the Hilbert space $\X:=L^2_{per}([0,2\pi],\R^2)$, the space of periodic square integrable functions on $[0,2\pi]$:
\bqq
\label{abstract}
\frac{dX}{dt}=\F(X,\mu),
\eqq
where 
$\F$ is the smooth nonlinear operator defined on $\X$ by
\bqq
\label{opF}
\F\left(\left(\begin{matrix} \psi(x)\\ \phi(x) \end{matrix} \right),\mu\right):= 
\left(
\begin{matrix}
- u (x) - (\mu+\alpha) v (x)+ \int_{-\pi}^\pi w(x-y) f(u(y))dy\\
\alpha\left(u(x)-v(x) \right)
\end{matrix}
\right).
\eqq
Note that with $U$ solution of \eqref{adbump}, $X_0=(U,U)$ is a stationary solution of \eqref{abstract} for all $\mu$. We define $\Lo_0\in \Lo(\X,\X)$ the bounded linear operator
\bqq
\label{opL0}
\Lo_0\left(\begin{matrix} \psi(x)\\ \phi(x) \end{matrix} \right):=
\left(\begin{matrix}
- \psi (x) - \alpha \phi (x) + \int_{- \pi}^{\pi} w(x - y) f'(U(y)) \psi (y) \d y \\
\alpha \psi(x) - \alpha \phi(x)
\end{matrix}
\right),
\eqq
such that $\Lo_0=D_X\F(X_0,0)$, the linearization of ${\mc F}$ right at the bifurcation point. 

Differentiating system \eqref{adbump} with respect to $x$ shows that $\zeta_0(x):=(U'(x),U'(x))$ is an eigenvector of the $\lambda=0$ eigenvalue of the linearized system \eqref{adlin} and thus of $\Lo_0$. Furthermore, if we denote $\zeta_1(x):=-(0,U'(x)/\alpha)$, then the following relations hold
\begin{align*}
\Lo_0 \zeta_1=\zeta_0,\ \ \ \ \ \ \ \ \  \Lo_0 \zeta_0 =0.
\end{align*}
We would like to conclude that $\lambda=0$ is an eigenvalue of $\Lo_0$ of algebraic multiplicity two and geometric multiplicity one. To this end, let us suppose that there exist two functions $\psi$ and $\phi$ such that
\bqs
\Lo_0\left(\begin{matrix} \psi\\ \phi\end{matrix} \right)=\zeta_1.
\eqs
The second component of this system reads
\bqs
-\alpha \phi = -\alpha \psi -\frac{U'}{\alpha},
\eqs
so that the first component can be written
\bqs
-(1+\alpha)\psi(x)+\int_{- \pi}^{\pi} w(x - y) f'(U(y)) \psi (y) \d y=\frac{U'}{\alpha}.
\eqs
Multiplying both sides of the above equation by $f'(U)U'$ and integrating over $[-\pi,\pi]$, we readily obtain
\begin{align*}
\frac{1}{\alpha}\|U'\|_f^2&=-(1+\alpha)\int_{-\pi}^\pi \psi(x)f'(U(x))U'(x)dx\\
&+\int_{-\pi}^\pi\left(\int_{- \pi}^{\pi} w(x - y) f'(U(y)) \psi (y) \d y\right)f'(U(x))U'(x)dx,
\end{align*}
with $\|U'\|_f^2:=\int_{-\pi}^\pi f'(U(x))U'(x)^2dx$. Inverting the order of integration in the double integral yields to the contradiction
\bqs
\frac{1}{\alpha}\|U'\|_f^2=0.
\eqs
As a consequence, $\lambda=0$ is an eigenvalue of $\Lo_0$ of algebraic multiplicity two and geometric multiplicity one.

Following the stability analysis of the previous section, we assume that the bump solution $U$ is such that 
\begin{align*}
\sigma_u&:=\sigma(\Lo_0)\cap \left\{ \lambda \in \C~;~ \Re(\lambda)>0 \right\} = \emptyset,\\
\sigma_c&:=\sigma(\Lo_0)\cap \left\{ \lambda \in \C~;~ \Re(\lambda)=0\right\}=\left\{ 0\right\},\\
\sigma_s&:= \sigma(\Lo_0)\cap \left\{ \lambda \in \C~;~ \Re(\lambda)<0 \right\} \subset \left\{  \lambda \in \C~;~ \Re(\lambda)<-\eta \right\},
\end{align*}
for some $\eta>0$ fixed. The last condition is always satisfied as the operator $\Lo_0$ is a compact operator on $\X$ and thus has discrete spectrum. It is also straightforward to verify that there exist $\omega_0>0$ and $C>0$ such that the following resolvent estimate is satisfied for each $|\omega|>\omega_0$
\bqs
\|(i\omega-\Lo_0)^{-1}\|_{\Lo(\X,\X)}\leq \frac{C}{|\omega|}.
\eqs
We denote $\caln:=\text{span}(\zeta_0,\zeta_1)$ and $P_c$ the spectral projection onto $\caln$ and define $P_s:=I-P_c$. The complementary space of $\caln$ in $\X$ is denoted $\cal S$ and $\mathcal{S}=P_s\X$. Following ideas developed by Iooss in \cite{iooss}, we can now apply a nonlinear center manifold type approach to our abstract system \eqref{abstract} and write any solution $X$ as
\bqq
\label{ansatz}
X=\mathcal{T}_{\Delta}\left(X_0+c \zeta_1 + \Psi(c,\mu) \right),
\eqq
with $\Psi(c,\mu)\in \mathcal{S}$, $\Psi(0,\mu)=0$ and $\partial_c\Psi(0,0)=0$. Here $\mathcal{T}_{\Delta}$ denotes the translation by $\Delta$:
\bqs
\mathcal{T}_{\Delta}(\psi(x),\phi(x))=(\psi(x+\Delta),\phi(x+\Delta)).
\eqs
Replacing the ansatz \eqref{ansatz} in \eqref{abstract}, we find immediately that
\bqq
\label{expansion}
\frac{d\Delta}{dt} \left( \zeta_0 +c \frac{d \zeta_1}{dx}+\frac{d \Psi(c,\mu) }{dx} \right)+\frac{d c}{dt}\zeta_1+\frac{dc}{dt}\partial_c\Psi(c,\mu)= \F(X_0+c \zeta_1 + \Psi(c,\mu),\mu).
\eqq
Here, we have used the translational invariance of our equations. Using normal form theory and the symmetries of the problem, we obtain the following amplitude equations for $\Delta$ and $c$
\begin{subequations}
\label{ampCM}
\begin{align}
\frac{d\Delta}{dt}&= c+\mathcal{O}\left(|c|(|\mu|+|c|^2)\right),\label{ampDelta}\\
\frac{dc}{dt}&=\Gamma_0\mu c +\Gamma_1 c^3+\mathcal{O}\left(|c|(|\mu|^2+|c|^4)\right).\label{ampc}
\end{align}
\end{subequations}
Now we determine the coefficients $\Gamma_0$ and $\Gamma_1$ that appear in equation \eqref{ampc}. We first need to Taylor expand $\F(X_0+Y,\mu)$ at $Y=0$ and we obtain
\bqs
\F(X_0+Y,\mu)=\Lo_0Y+\mu \calr_1 (Y)+\calr_2(Y,Y)+\calr_3(Y,Y,Y)+\text{h.o.t}
\eqs
where
\bqs
\calr_1 \left(\left(\begin{matrix} y_1\\ y_2 \end{matrix} \right) \right)=  \left(\begin{matrix} -y_2\\ 0 \end{matrix} \right),
\eqs

\bqs
\calr_2 \left(\left(\begin{matrix} y_1\\ y_2\end{matrix} \right),\left(\begin{matrix} z_1\\ z_2 \end{matrix} \right) \right)(x)= \left(\begin{matrix} \int_{-\pi}^\pi w(x-y) f_2(y) y_1(y)z_1(y)dy\\ 0 \end{matrix} \right),
\eqs

and 

\bqs
\calr_3 \left(\left(\begin{matrix} x_1\\ x_2 \end{matrix} \right),\left(\begin{matrix} y_1\\ y_2 \end{matrix} \right),\left(\begin{matrix} z_1\\ z_2 \end{matrix} \right) \right)(x)=  \left(\begin{matrix} \int_{-\pi}^\pi w(x-y)f_3(y) x_1(y) y_1(y)z_1(y)dy\\ 0 \end{matrix} \right)
\eqs
with $f_k(x)=f^{(k)}(U(x))/(k!)$. We also Taylor expand $\Psi$ at $(0,0)$:
\bqs
\Psi(c,\mu)=\mu c \Psi_{11}+c^2\Psi_{20}+c^3\Psi_{30}+\text{h.o.t.}
\eqs
We will also need to compute the adjoint operator $\Lo_0^*$ of $\Lo_0$ defined by
\bqq
\label{opL0adj}
\Lo_0^*\left(\begin{matrix} \psi(x)\\ \phi(x) \end{matrix} \right):=
\left(\begin{matrix}
- \psi (x) + \alpha \phi (x) + f'(U(x))\int_{- \pi}^{\pi} w(x - y) \psi (y) \d y \\
-\alpha \psi(x) - \alpha \phi(x)
\end{matrix}
\right).
\eqq
Note that if we define 
\bqs
\zeta_0^*(x)= \frac{f'(U(x))U'(x)}{\|U'\|_f^2} \left(\begin{matrix}1 \\ 0 \end{matrix} \right)\text{ and } \zeta_1^*(x)= \frac{f'(U(x))U'(x)}{\|U'\|_f^2}  \left(\begin{matrix}\alpha \\ -\alpha \end{matrix} \right)
\eqs
with $\|U'\|_f^2:=\int_{-\pi}^\pi f'(U(x))U'(x)^2dx$, then we have the relations
\begin{align*}
\Lo_0^* \zeta_0^*&=\zeta_1^* \text{ and }\Lo_0^*\zeta_1^*=0,\\
\langle \zeta_0,\zeta_0^* \rangle &= 1 \text{ and } \langle \zeta_0,\zeta_1^* \rangle=0\\
\langle \zeta_1,\zeta_1^* \rangle &= 1 \text{ and } \langle \zeta_1,\zeta_0^* \rangle=0.
\end{align*}
Collecting the $\mathcal{O}(\mu c)$ terms in the expansion \eqref{expansion} we obtain
\bqs
\Gamma_0 \zeta_1 = \Lo_0 \Psi_{11}+\calr_1(\zeta_1)
\eqs
which gives
\bqq
\label{Gamma0}
\Gamma_0=\langle \Psi_{11},\Lo_0^*\zeta_1^*\rangle_\X+ \langle \calr_1(\zeta_1) , \zeta_1^*\rangle_\X=1.
\eqq
Collecting the $\mathcal{O}(c^3)$ terms in the expansion \eqref{expansion} we obtain
\bqs
\frac{d\Psi_{20}}{dx}+\Gamma_1 \zeta_1=\Lo_0\Psi_{30}+2\calr_2(\zeta_1,\Psi_{20})+\calr_3(\zeta_1,\zeta_1,\zeta_1).
\eqs
Note that $\calr_2(\zeta_1,\Psi_{20})=\calr_3(\zeta_1,\zeta_1,\zeta_1)=0$ such that the above equation projected on $\zeta_1$ gives
\bqs
\Gamma_1=-\left\langle \frac{d\Psi_{20}}{dx},\zeta_1^*\right\rangle_\X.
\eqs
The equation for $\Psi_{20}$ is found by collecting $\mathcal{O}(c^2)$ terms in the expansion \eqref{expansion} yielding
\bqs
\frac{d\zeta_1}{dx}=\Lo_0\Psi_{20}+\calr_2(\zeta_1,\zeta_1).
\eqs
Using the fact that $\calr_2(\zeta_1,\zeta_1)=0$, we find an expression for $\Psi_{20}$ of the form
\bqs
\Psi_{20}=\delta \zeta_0+\frac{U''}{\alpha^2}\left(\begin{matrix}0 \\ 1 \end{matrix} \right)+\varphi \left( \begin{matrix} 1 \\ 1 \end{matrix} \right)
\eqs
where $\delta\in\R$ is a constant and $\varphi$ is a solution of
\bqs
-(1+\alpha)\varphi(x)+\int_{-\pi}^\pi w(x-y)f'(U(y))\varphi(y)dy=\frac{U''(x)}{\alpha}.
\eqs
Finally, we obtain
\begin{align*}
-\left\langle \frac{d\Psi_{20}}{dx},\zeta_1^*\right\rangle_\X&=-\delta \left\langle \frac{d\zeta_0}{dx},\zeta_1^*\right\rangle_\X-\frac{1}{\alpha^2}\left\langle U'''\left( \begin{matrix} 0 \\ 1 \end{matrix} \right),\zeta_1^*\right\rangle_\X-\left\langle \frac{d\varphi}{dx}\left( \begin{matrix} 1 \\ 1 \end{matrix} \right),\zeta_1^*\right\rangle_\X\\
&=\frac{\langle U''',U' \rangle_f}{\alpha \|U'\|_f^2},
\end{align*}
with $\langle U''',U' \rangle_f=\int_{-\pi}^\pi f'(U(x))U'''(x)U'(x)dx$.
This gives
\bqq
\label{Gamma1}
\Gamma_1=\frac{\langle U''',U' \rangle_f}{\alpha \|U'\|_f^2}.
\eqq
In the specific case of an even, monotonically decreasing connectivity function $w$, such as $w(x)=\cos(x)$, it is possible to determine the sign of $\Gamma_1$ using the identity
\bqs
(1+\alpha)U''(x)=\int_{-\pi}^\pi w(x-y)\frac{d^2f(U(y))}{dy^2}dy.
\eqs
Indeed, in that case, we have
\begin{align*}
\langle U''',U' \rangle_f&=\int_{-\pi}^\pi f'(U(x))U'''(x)U'(x)dx = \int_{-\pi}^\pi \frac{d f(U(x))}{dx} U'''(x)dx \\
&= - \int_{-\pi}^\pi \frac{d^2 f(U(x))}{dx^2} U''(x)dx \\
&= -\frac{1}{1+\alpha}\int_{-\pi}^\pi \int_{-\pi}^\pi w(x-y) \frac{d^2 f(U(x))}{dx^2} \frac{d^2 f(U(y))}{dy^2}dxdy <0.
\end{align*}
Thus, it is clear from \eqref{ampc} that we recover the pitchfork bifurcation discussed in the previous section. In particular, for $\alpha < \beta$, there are three constant speed solutions of \eqref{ampc}, corresponding to an unstable stationary bump and a pair of stable counter-propagating pulses with speeds\footnote{Note for the cosine weight function (\ref{cos}), we have $U = A \cos x$, so $||U'||^2_f = |\langle U''',U' \rangle|$ and $c_{\pm} = \pm \sqrt{\alpha ( \beta - \alpha)}$. As we will show in section \ref{existpulse}, this approximation happens to be exact.} 
\bqq
\label{speeds}
c_0=0,\quad \quad \ \ \ \ \ \ \   c_\pm=\pm \sqrt{\alpha (\beta-\alpha)\frac{\|U'\|_f^2}{\left| \langle U''',U' \rangle_f \right|}}.
\eqq  
Using equation \eqref{ansatz}, we find 
\begin{align*}
u_\pm(x,t)&=  U(x+c_\pm t)+\mathcal{O}(\mu), \\
v_\pm(x,t)&=  U(x+c_\pm t)-\frac{c_\pm}{\alpha}U'(x+c_\pm t) + \mathcal{O}(\mu).
\end{align*}
Close to the bifurcation point the shape of the propagating pulses is approximately the same as the stationary bump, except that the recovery variable is shifted relative to $u$ by an amount proportional to the speed $c_\pm$, that is,
\bqs
u_\pm(x,t)\approx U(\xi), \quad v_\pm(x,t)\approx U(\xi-c_\pm/\alpha), \quad \xi=x+c_\pm t.
\eqs
An analogous result on drift instability of fronts was previously obtained for both reaction-diffusion equations \cite{hagberg94,bode97} and neural field equations \cite{bressloff03}.

\section{Perturbed amplitude equations at the drift bifurcation}  \label{stochcm}
Now, we study how the amplitude equations \eqref{ampCM} are perturbed when we introduce additive noise in the neural field equations \eqref{ringad}. Throughout this section, we assume that $\mu=\beta-\alpha>0$ is small. Following ideas developed for the Swift-Hohenberg equations \cite{blomker01,blomker03,blomker05}, we consider only random perturbations $\epsilon \d W(x,t)$ for which $\epsilon = \mu^2 \chi$, with $\chi>0$. From the form of the amplitude equation for $c$ given in equation \eqref{ampc}, we readily see that $c(t)\sim \sqrt{\mu}~C(\mu t)$ as $\mu\rightarrow 0$, where $C$ satisfies the cubic equation
\begin{align*}
\label{cubic}
C'=C +\Gamma_1 C^3.
\end{align*}
Thus, bumps drift slowly when close to the instability, so $\Delta(t)\sim \sqrt{\mu}t$. 

To perform a perturbation expansion in $\mu$, we introduce two different time scales $\tau=\sqrt{\mu}t$ and $\hat \tau = \mu t$. Then, we rewrite system \eqref{radlang} in the more convenient form
\begin{subequations} \label{slow}
\begin{align}
\sqrt{\mu}\d u (x,\tau) &= \left[ - u (x,\tau) - \beta v (x,\tau) + \int_{- \pi}^{\pi} w(x-y) f(u(y,\tau)) \d y \right] \d \tau,  \\
\sqrt{\mu} \d v (x,\tau) &= \alpha \left[ u (x,\tau) - v (x,\tau) \right] \d \tau + \mu^{\frac{3}{2}} \chi \d \widehat{W} (x,\hat\tau),
\end{align}
\end{subequations}
where $\d \widehat{W} (x,\hat\tau):=\sqrt{\mu}\d W(x,\mu^{-1}\hat\tau)$ is a rescaled version of the Wiener process $\d W$ that is independent of the parameter $\mu$. Here, we   have used the scaling properties of the Brownian motion: the processes $\left(W(x,t) \right)_{t\geq0}$ and $\left(\mu^{-\frac{1}{2}}W(x,\mu t) \right)_{t\geq0}$ are in law the same process. Finally, we  look for solutions $X$ of \eqref{slow} that can be expanded in the form
\begin{subequations} \label{ansatzSto}
\begin{align}
u(x,\tau)&=U(x-\Delta(\tau))+\mu u_1(x-\Delta(\tau),\tau)+\mu^{\frac{3}{2}}u_2(x-\Delta(\tau),\tau)+\mathcal{O}(\mu^2),\\
v(x,\tau)&=U(x-\Delta(\tau))-\sqrt{\mu}\frac{C(\hat \tau )}{\alpha}U'(x-\Delta(\tau))+\mu v_1(x-\Delta(\tau),\tau)\nonumber\\
&~+\mu^{\frac{3}{2}}v_2(x-\Delta(\tau),\tau) +\mathcal{O}(\mu^2).
\end{align}
\end{subequations}
Collecting terms at order $\mathcal{O}(\sqrt{\mu})$ we find that
\begin{align*}
\d \Delta = C d\tau.
\end{align*}
At order $\mathcal{O}(\mu)$ we obtain the system
\bqq
\label{eq:mu1}
\Lo_0\left(\begin{matrix} u_1 \\ v_1\end{matrix} \right)=\frac{C^2}{\alpha}\left(\begin{matrix} 0 \\ U'' \end{matrix} \right).
\eqq
Using the Ito formula on the last component of \eqref{eq:mu1} yields the expansion
\bqq
\label{eq:resultint}
\d\left[ u_1(x-\Delta(\tau),\tau)-v_1(x-\Delta(\tau),\tau) \right]=-\frac{C^3}{\alpha^2}U'''(x-\Delta(\tau))d\tau+\mathcal{O}\left(\mu^{\frac{1}{2}}\right).
\eqq
At order $\mathcal{O}(\mu^{\frac{3}{2}})$ we have
\begin{subequations} \label{eq:muint}
\begin{align}
\d \left[u_1(x-\Delta(\tau),\tau)\right]&=\left[\mathcal{M}\left(u_2(x-\Delta(\tau),\tau)\right)+\frac{U'(x-\Delta(\tau))}{\alpha}C\right]d\hat \tau\\
\d \left[v_1(x-\Delta(\tau),\tau)\right]&=\frac{U'(x-\Delta(\tau))}{\alpha}\d C + \alpha(u_2(x-\Delta(\tau),\tau)-v_2(x-\Delta(\tau),\tau))\nonumber\\
&~+\chi \d \widehat{W} (x,\hat\tau),
\end{align}
\end{subequations}
where
\bqs
\mathcal{M}u(x)=-(1+\alpha)u(x)+\int_{-\pi}^\pi w(x-x')f'(U(x'))u(x')dx'.
\eqs
 Combining equation \eqref{eq:resultint} with system \eqref{eq:muint}, we obtain the equation
 \bqs
\left[ \mathcal{M} u_2\right] \d \hat \tau= \frac{U'}{\alpha}\d C+ \left[-\frac{U'}{\alpha}C-\frac{U'''}{\alpha^2}C^3 \right]\d\hat \tau -\chi \d \widehat{W} (x,\hat\tau).
 \eqs
The above equation can further be projected along $\alpha f'(U)U'$, yielding the stochastic amplitude equation
\bqs
\d C=\left[C+\Gamma_1C^3\right]\d \hat \tau-\chi\alpha \d \mathcal{W}(\hat \tau),
\eqs
where the process $\d \mathcal{W}(\hat \tau)$ is defined as
\bqs
\d \mathcal{W}(\hat \tau)=\frac{\langle \d \widehat{W} (x,\hat\tau),U'\rangle_f}{\|U'\|^2_f}.
\eqs
Due to our careful choice of $\epsilon = \mu^2 \chi$, the amplitude equation obtained for $C$ is independent of the bifurcation parameter $\mu$. This equation is called the \textit{stochastically forced Landau equation} in \cite{blomker01,blomker03,blomker05}.
Summarizing, we have obtained the following stochastic system satisfied by $\Delta$ and $C$ from our ansatz \eqref{ansatzSto}
\begin{subequations}
\label{ampSto}
\begin{align}
\d\Delta &= C\d \tau \\
\d C&=\left[C+\Gamma_1C^3\right]\d \hat \tau-\chi\alpha \d \mathcal{W}(\hat \tau).
\end{align}
\end{subequations}

\begin{figure}
\begin{center}  \includegraphics[width=13cm]{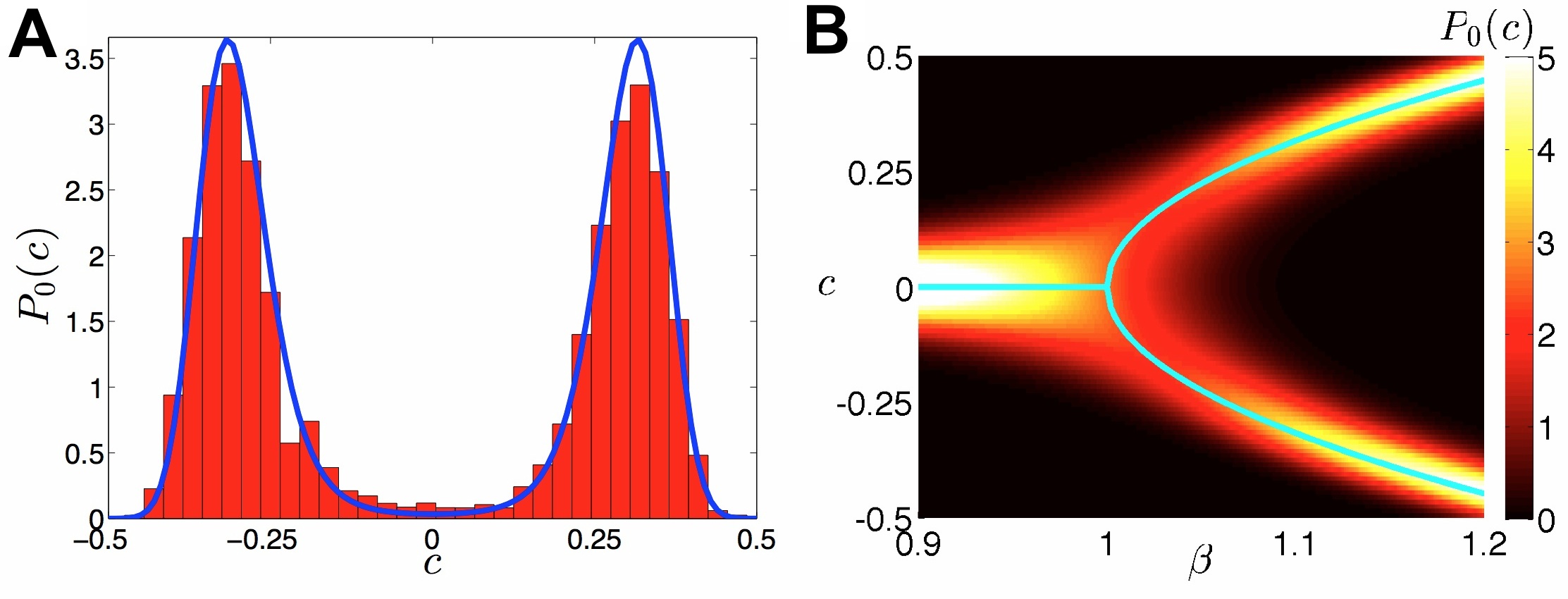} \end{center}
\caption{Stationary density $P_0(c)$ of the speed $c$ of the traveling pulse arising through a pitchfork bifurcation for sufficiently strong amplitude ($\beta > \alpha$) linear adaptation in (\ref{radlang}). Here we take the cosine weight (\ref{cos}), Heaviside firing rate function (\ref{H}), and cosine spatial noise correlation ${\bf C}(x) = \cos (x)$. {\rm ({\bf A})} Comparing numerically computed histogram (red bars) to theoretical prediction (\ref{statdist}) (blue curve) of the stationary density $P_0(c)$ for $\beta = 1.1$. The numerically computed histogram uses a realization of (\ref{radlang}) lasting $10^5$ time units solved with Euler-Maruyma with timestep $dt = 0.01$. {\rm ({\bf B})} Stationary density $P_0 (c)$ (\ref{statdist}) plotted as $\beta$ is varied, showing how the pitchfork bifurcation at $\beta = \alpha = 1$ affects the pulse's speed $c$. Other parameters are $\theta = 0.25$ and $\ve = 0.03$.}
\label{stochpitch}
\end{figure}

Since $\Delta$ does not appear in (\ref{ampSto}b), we can analyze the long term behavior of $C$ independently from $\Delta$. First, to link these results back to the original stochastic system (\ref{radlang}), we rescale time ($c(t) \sim \sqrt{\mu} C( \mu t)$) so that we obtain the stochastic differential equation
\begin{align}
\d c (t) = \left[ ( \beta - \alpha) c (t) + \Gamma_1 c^3 (t) \right] \d t + \d B (t)  \label{cstamp}
\end{align}
where $\d B (t)$ is the increment of a white noise process so $\langle B(t) \rangle = 0$ and $\langle B(t)^2 \rangle = {\bf D} t$ with
\begin{align}
{\bf D} = \ve^2 \alpha^2 \frac{\D \int_{- \pi}^{\pi} \int_{- \pi}^{\pi} {\bf C}(x-y)f'(U(y))U'(y) f'(U(x)) U'(x) \d x \d y}{\D \left[ \int_{- \pi}^{\pi} f'(U(x)) U'(x)^2 \d x \right]^2}. \label{coeffDiff}
\end{align}
The deterministic dynamics associated with (\ref{cstamp}) moves along the gradient of the potential function
\begin{align}
{\bf U}(c) = - \frac{\beta - \alpha}{2} c^2 - \frac{\Gamma_1}{4} c^4.  \label{cpotwell}
\end{align}
Thus, (\ref{ampSto}b) can be reformulated as an equivalent Fokker-Planck equation \cite{gardiner04}
\begin{align}
\frac{\pd P ( c, t)}{\pd t} = \frac{\pd}{\pd c} \left[ {\bf U}'(c) P(c,t) \right] + \frac{{\bf D}}{2} \frac{\pd^2 P(c,t)}{\pd c^2},  \label{campfp}
\end{align}
where $P(c,t)$ is the probability density of observing speed $c$ at time $t$. The stationary solution to (\ref{campfp}) is then given
\begin{align}
P_0 ( c) = {\mc N} \e^{- 2 {\bf U} (c)/ {\bf D}},  \label{statdist}
\end{align}
where ${\mc N}^{-1} = \int_{- \infty}^{\infty} \e^{- 2 {\bf U}(c) / {\bf D}} \d c$ is a normalization factor. To demonstrate this analysis, we consider the specific case where the weight function is a cosine (\ref{cos}), the firing rate function is a Heaviside (\ref{H}), and the spatial correlations in noise are given by the cosine ${\bf C}(x) = \cos (x)$. In this case, $U(x) = A \cos (x)$ with $A$ given in \eqref{ampAheaviside} so we can compute $\Gamma_1 = \alpha^{-1}$ and  the diffusion coefficient:
\bqs
{\bf D}=\frac{\ve^2 \alpha^2 (1+\beta)^2}{2 + 2 \sqrt{1 - (1+ \beta)^2 \theta^2}}.
\eqs
We demonstrate the accuracy of the stochastic amplitude equation (\ref{cstamp}) for $c(t)$ by computing the stationary distribution $P_0 (c)$ from numerical simulations and comparing it with our theoretically derived formula (\ref{statdist}) in Fig. \ref{stochpitch}({\bf A}). Specifically, we take the cosine weight (\ref{cos}), Heaviside firing rate (\ref{H}), and cosine spatial correlations ${\bf C}(x) = \cos x$. In addition, we demonstrate how the stationary density (\ref{statdist}) varies with adaptation strength $\beta$, splitting from a unimodal to bimodal distribution as $\beta$ passes through the pitchfork bifurcation at $\alpha$ (Fig. \ref{stochpitch}({\bf B})). Previously, Laing and Longtin studied the effects of noise upon the propagation of traveling pulses in a network where adaptation nonlinearly affected neural activity \cite{laing01b}. Specifically, they found that noise could shift the location of the pitchfork bifurcation that generated traveling pulses from destabilized stationary bumps. Here, we find no such shift, perhaps due to the purely linear effect adaptation has upon the neural activity variable in (\ref{radlang}). We also note that we have extended the work in \cite{laing01b} by providing a principled derivation of a stochastic amplitude equation for the noisy pitchfork bifurcation, which can then be used to analytically compute statistics of traveling pulse speed.

\begin{figure}
\begin{center} \includegraphics[width=13cm]{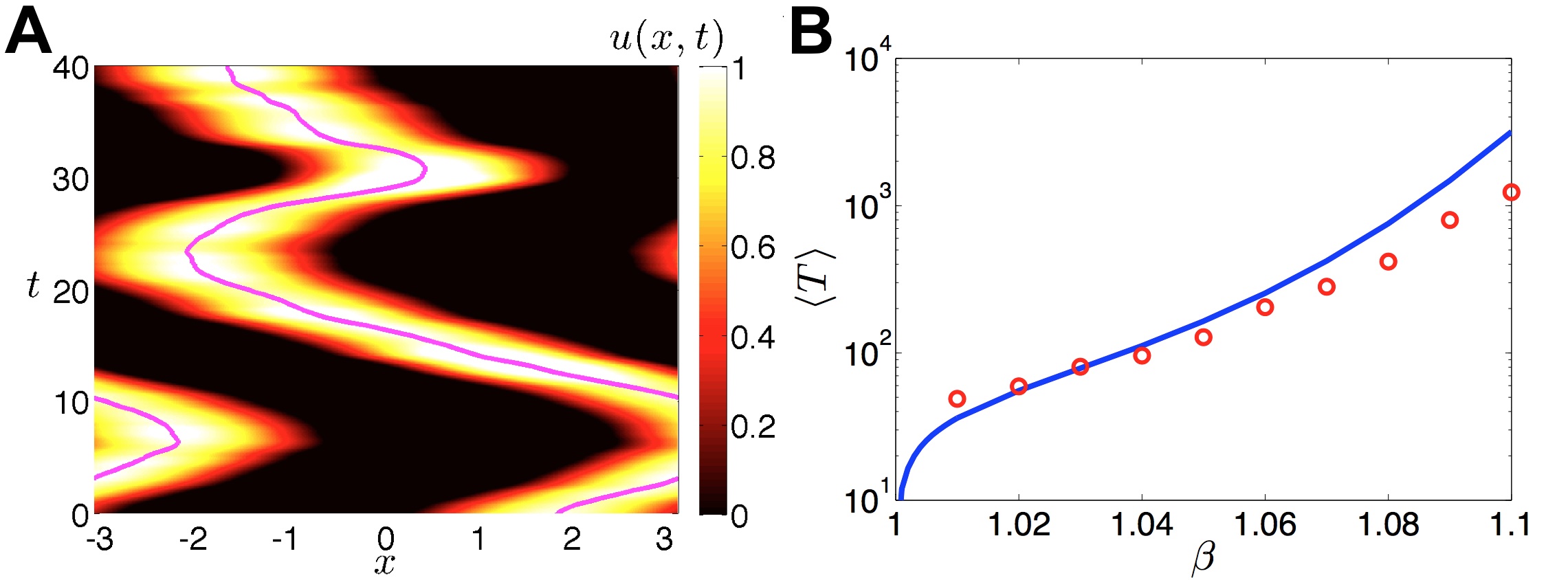} \end{center}
\caption{Noise-induced switching in the direction of propagation of traveling pulses in the vicinity of the pitchfork bifurcation at $\beta = \alpha$. Again, we take the cosine weight (\ref{cos}), Heaviside firing rate function (\ref{H}), and cosine spatial noise correlation ${\bf C}(x) = \cos (x)$. {\rm ({\bf A})} Space-time plot of a traveling pulse for $\beta = 1.05$ that intermittently switches its direction of propagation when forced by sufficiently large noise kicks. {\rm ({\bf B})} The average amount of time $\langle T \rangle$ between propagation direction switches plotted as a function of $\beta$ calculated theoretically (blue line) using (\ref{avswtime}) and numerically (red dots) using a realization of (\ref{radlang}) lasting $10^5$ time units solved with Euler-Maruyma with timestep $dt = 0.01$. Other parameters $\alpha = 1$, $\theta = 0.25$, and $\ve = 0.03$.}
\label{pulseswitch}
\end{figure}

Lastly, we explore propagation direction switching brought about by fluctuations (as in Fig. \ref{pulseswitch}({\bf A})). The speed of the traveling pulse $c(t)$ fluctuates about either $c_{\pm}$ for $\beta > \alpha$, but eventually fluctuations force the system substantially that the nearby speed switches from $c_{\pm}$ to $c_{\mp}$. Essentially, this occurs through a large deviation in noise that moves the dynamics from one of the two attractors defined by the double-well potential ${\bf U}(c)$ to the other. The frequency with which these transitions occur can be characterized by computing the mean first passage time of the variable $c(t)$ through the potential barrier at $c \equiv 0$. Since the system (\ref{cstamp}) is symmetric about zero, we can consider the case where $c(t)$ moves from near $c_-$ to near $c_+$, giving us the formula \cite{gardiner04}
\begin{align}
\langle T \rangle = \frac{2}{{\bf D}} \int_{c_-}^0 \e^{2 {\bf U}(y) / {\bf D}} \int_{- \infty}^y  \e^{-2 {\bf U}(z) / {\bf D}} \d z \d y.  \label{avswtime}
\end{align}
The formula predicts the average time to a transition between propagation directions ($\pm$) fairly well, as demonstrated in Fig. \ref{pulseswitch}({\bf B}). As the adaptation strength $\beta$ increases away from the adaptation rate $\alpha$ value, the systems dynamics moves farther from the pitchfork bifurcation. Thus, the depth of either potential well in (\ref{cpotwell}) decreases, so it takes longer until a transition. As would be expected, the average length of time until a switch $\langle T \rangle$ scales roughly exponentially with the bifurcation parameter $\beta$.

%


\section{Existence of traveling pulses}\label{existpulse}

Well beyond the pitchfork bifurcation (outside where $|\beta - \alpha|^2 \sim {\mc O}( \ve )$), the dynamics of the system (\ref{ringad}) are best studied by constructing the resulting traveling pulse solutions. Doing so allows us to study how the speed varies as a function of $\alpha$ and $\beta$. We can also examine the linear stability of these pulses, which should capture similar findings to our amplitude equation calculations.  It should be noted that there are a few related studies of this bifurcation on the infinite domain and the plane \cite{laing01b,pinto01,folias05,coombes12}, which exploited a Heaviside firing rate function (\ref{H}). However, we recapitulate these findings here in the ring network in the general setting of a sigmoid firing rate function (\ref{sig}).

To start, we look for traveling wave solutions of (\ref{ringad}), such that $(u,v) = (U( \xi) , V( \xi ))$ where $\xi = x - ct$ is a wave coordinate with associated wave speed $c$. Under this constraint, solutions to (\ref{ringad}) are given by the system
\begin{subequations} \label{adpulsys}
\begin{align}
- c U'( \xi ) &= - U( \xi ) - \beta V( \xi ) + \int_{- \pi}^{\pi} w( \xi - y ) f( U(y)) \d y,  \label{adpulsys1} \\
- c V'( \xi ) &= \alpha U ( \xi ) - \alpha V( \xi ).  \label{adpulsys2}
\end{align}
\end{subequations}
Equivalently, we can consider traveling pulse solutions to the second-order equation (\ref{secordeq}) given by \cite{folias04,kilpatrick08}

\begin{align}
 \mathcal{Q}U(\xi)= \int_{- \pi}^{\pi} w( \xi - y ) \left[ -c f'( U(y)) U'(y) + \alpha f(U(y)) \right] \d y, \label{adpul2nd}
\end{align}
where $\mathcal{Q}$ is now given by
\begin{align}
 \mathcal{Q}U(\xi)=c^2 U''( \xi ) - c ( 1+ \alpha ) U'( \xi ) + \alpha ( 1+ \beta ) U( \xi ).
\end{align}
We begin by briefly explaining a general procedure for constructing traveling pulse solutions with arbitrary firing rate and weight functions in the system (\ref{adpul2nd}). To begin, we expand $U( \xi )$ in a Fourier series
\begin{align}
U( \xi ) = \sum_{k=0}^{N} A_k \cos (k \xi ) + \sum_{k=1}^N B_k \sin (k \xi )  \label{Utpexp}
\end{align}
where $N$ is the maximal mode needed to fully characterize $w(x)$ given by (\ref{wfexp}), as in our analysis of stationary bump solutions. By plugging (\ref{Utpexp}) and (\ref{wfexp}) into (\ref{adpul2nd}), we can generate a system of nonlinear equations for the coefficients
\begin{align}
c^2 k^2 A_k + c k (1 + \alpha ) B_k - \alpha ( 1+ \beta ) A_k &= w_k \int_{- \pi}^{\pi} \cos (k x) [c f'(U(x))U'(x) - \alpha f(U(x))] \d x, \label{AkBkpulse} \\
c^2 k^2 B_k  - ck (1+ \alpha ) A_k - \alpha (1 + \beta ) B_k &= w_k \int_{- \pi}^{\pi} \sin (kx) [c f'(U(x)) U'(x) - \alpha f(U(x))] \d x. \nonumber
\end{align}
In general, we could then use numerical methods to solve for all of the coefficients $A_k, B_k$ for $k=1,...,N$. We now demonstrate our ability to generate explicit solutions analytically for the case of the cosine weight (\ref{cos}) so that $U(\xi ) = A \cos ( \xi ) + B \sin ( \xi )$ and the system (\ref{AkBkpulse}) is given
\begin{align*}
[c^2 - \alpha (1 + \beta)] A + c (1 + \alpha ) B  &= \int_{- \pi}^{\pi} \cos (x) [c f'(U(x))U'(x) - \alpha f(U(x))] \d x, \\
- c (1+ \alpha ) A  + [c^2 - \alpha (1 + \beta )] B &= \int_{- \pi}^{\pi} \sin (x) [c f'(U(x)) U'(x) - \alpha f(U(x))] \d x,
\end{align*}
which can be solved for
\begin{subequations}  \label{adpulsAB}
\begin{align}
A &= \frac{(c^2 - \alpha ( 1+ \beta )) {\mc C} - c (1+ \alpha) {\mc S}}{(c^2 - \alpha ( 1+ \beta ))^2 + c^2 (1+ \alpha )^2},  \\
B &= \frac{c(1+ \alpha) {\mc C} + ( c^2 - \alpha (1 + \beta )) {\mc S}}{(c^2 - \alpha ( 1+ \beta ))^2 + c^2 (1+ \alpha )^2},
\end{align}
\end{subequations}
where
\begin{subequations} \label{adpulcs}
\begin{align}
{\mc C} &=  \int_{- \pi}^{\pi} \cos x \left[ c f'( U(x)) U'(x) - \alpha f(U(x)) \right] \d x, \\
{\mc S} &=  \int_{- \pi}^{\pi} \sin x \left[ c f'(U(x) U'(x) - \alpha f(U(x))) \right] \d x. 
\end{align}
\end{subequations}
Thus, we need to find the roots of the nonlinear system generated by plugging the expressions (\ref{adpulsAB}) into (\ref{adpulcs}) to yield
\begin{align*}
{\mc C} &= - A c {\mc J} ( \sin x \cos x ) - A \alpha {\mc J} ( \sin^2 x) + B c {\mc J} ( \cos^2 x ) + B \alpha {\mc J} ( \cos x \sin x ), \\
{\mc S} &= - A c {\mc J} ( \sin^2 x ) + A \alpha {\mc J} ( \cos x \sin x ) + B c {\mc J} ( \cos x \sin x ) - B \alpha {\mc J} ( \cos^2 x ),
\end{align*}
where
\begin{align*}
{\mc J} (r(x)) = \int_{- \pi}^{\pi} r(x) f'(A \cos x + B \sin x ) \d x.
\end{align*}
Due to the translation symmetry of (\ref{ringad}), there will be a continuum of solutions $(A,B)$. Considering a monotone increasing firing rate function such as the sigmoid (\ref{sig}), we can break the degeneracy of solutions by fixing the pulse's position by requiring the leading edge of $U$ cross above a threshold $\theta$ at $\xi = \pi$. This provides us with two implicit equations for the width $a$ and speed $c$ of the pulse
\begin{align*}
U( \pi ) = - A = \theta, \ \ \ \ \ \ \ \ \ \ \ \ \ U( \pi - a) = A \cos ( \pi -a ) + B \sin ( \pi - a) = \theta.
\end{align*}
However, we can also generate explicit solutions in the case of a Heaviside firing rate function (\ref{H}). First, we can compute the constants ${\mc C}$ and ${\mc S}$ using the formulae (\ref{adpulcs}) as well as noting that $U( \xi ) > \theta$ when $x \in ( \pi - a, \pi )$ and $U( \xi ) < \theta$ otherwise. This yields
\begin{align*}
{\mc C} = \alpha \sin a + c ( 1- \cos a), \ \ \ \ \ \ \ \ \ \ {\mc S} = c \sin a - \alpha ( 1- \cos a).
\end{align*}
Plugging this into our formulae for the prefactors $A$ and $B$ given by (\ref{adpulsAB}) and imposing the threshold conditions $U( \pi ) = U ( \pi - a) = \theta$, we have
\begin{subequations}
\begin{align}
\frac{(c^2 + \alpha^2 (1 + \beta) ) \sin a - ( c^3 + c \alpha^2 - c \alpha \beta )( 1- \cos a)}{(c^2  - \alpha (1 + \beta ))^2 + c^2 ( 1+ \alpha )^2} &= \theta ,\label{adHthresh1} \\
\frac{(c^2 + \alpha^2(1 + \beta) ) \sin a + ( c^3 + c \alpha^2 - c \alpha \beta )( 1- \cos a )}{(c^2  - \alpha (1 + \beta ))^2 + c^2 ( 1+ \alpha )^2} &= \theta.  \label{adHthresh2}
\end{align}
\end{subequations}
By taking the difference of (\ref{adHthresh2}) and (\ref{adHthresh1}), we can generate the equation
\begin{align*}
\frac{c^3 + c \alpha^2 - c \alpha \beta}{(c^2 - \alpha ( 1 + \beta ))^2 + c^2 ( 1+ \alpha)^2} ( 1- \cos a ) = 0,
\end{align*}
since $a=0$ is a trivial solution, we study solutions where
\begin{align*}
c^3 +  c \alpha^2 - c \alpha \beta = 0.
\end{align*}
\begin{figure}
\begin{center} \includegraphics[width=12.4cm]{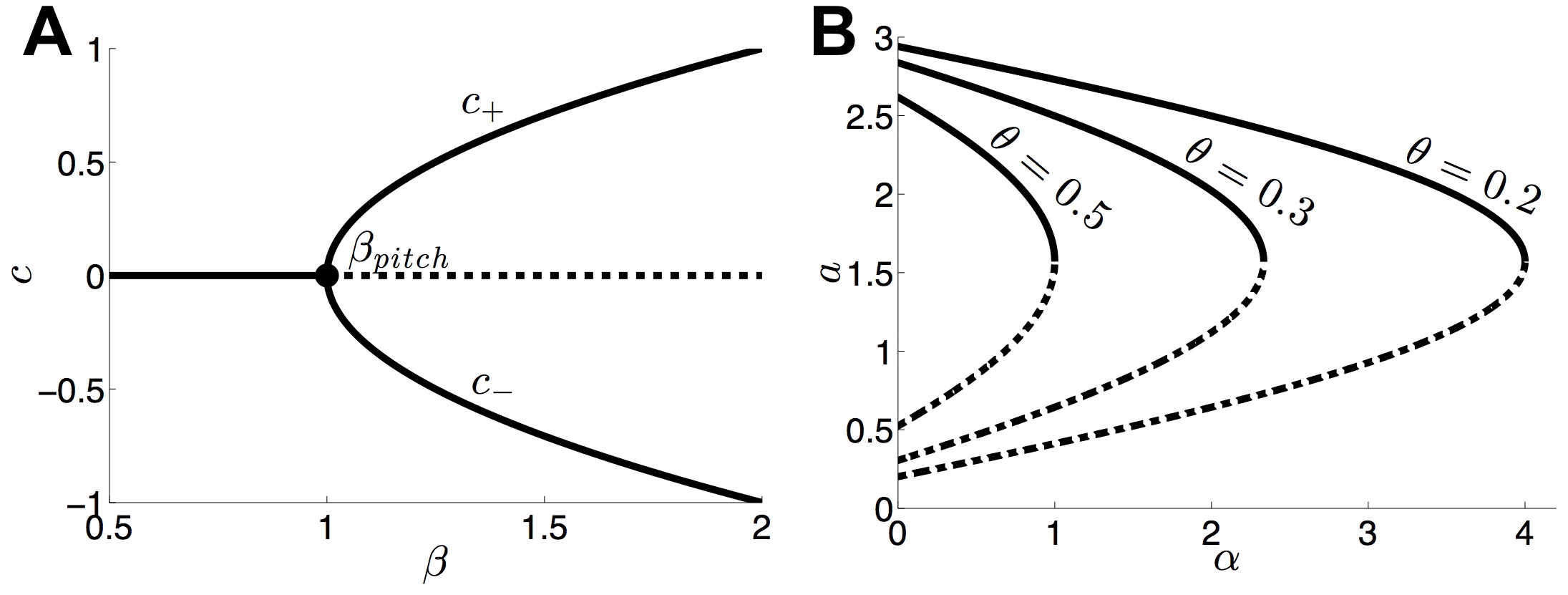}  \end{center}
\caption{{\rm ({\bf A})} Traveling pulse speed $c$ as a function of adaptation strength $\beta$. Speed increases with $\beta$ past the pitchfork bifurcation point $\beta_{pitch}$. {\rm ({\bf B})} Width a of stable (solid) and unstable (dashed) traveling pulses as a function of adaptation rate $\alpha$.}
\label{expulplot}
\end{figure}
Thus, we have a cubic providing us up to three possible speeds for a traveling pulse solution. The trivial $c=0$ solution, we would expect, as it is the limiting case of stationary bump solutions that we have already studied, which will be unstable when $\beta >\alpha$. In line with our previous findings, for $\beta \geq \alpha$, we have the two additional solutions
\begin{align}
c_{\pm} = \pm \sqrt{ \alpha (\beta - \alpha)}   \label{adpulspeed}
\end{align}
so that either provides a right-moving ($+$) and left-moving ($-$) traveling pulse solution. Thus, as discussed, when $\beta$ is beyond the pitchfork bifurcation point $\beta_{pitch} = \alpha$, there emerge two stable traveling pulse solutions. The pulse widths are then given by taking the mean of (\ref{adHthresh1}) and (\ref{adHthresh2}) to give
\begin{align}
\frac{c^2  + \alpha^2(1 + \beta)}{(c^2 - \alpha(1 + \beta ))^2 + c^2 ( 1 + \alpha )^2} \sin a = \theta.  \label{adpultsum}
\end{align}
Upon plugging the expression for the wavespeed (\ref{adpulspeed}) into (\ref{adpultsum}) and simplifying, we find
\begin{align}
\frac{\sin a}{1 + \alpha} = \theta  \label{sinathet}
\end{align}
meaning we can expect to find four traveling pulse solutions, two with each speed (\ref{adpulspeed}) that have widths
\begin{align}
a_s = \pi - \sin^{-1} [ \theta ( 1 + \alpha)], \ \ \ \  \ a_u = \sin^{-1} [ \theta ( 1 + \alpha ) ].  \label{adpulwid}
\end{align}
We can find, using linear stability analysis, that the two traveling pulses associated with the width $a_s$ are stable. Plugging the speed (\ref{adpulspeed}) and width-threshold relationship (\ref{sinathet}) into the formulae for the coefficients $A$ and $B$, we find
\begin{align}
U( \xi ) =  \frac{1 - \cos a}{1+ \alpha} \sin \xi - \frac{\sin a}{1 + \alpha} \cos \xi.  \label{Upulsol}
\end{align}
Subsequently solving equation (\ref{adpulsys2}), we find
\begin{align}
V( \xi ) = \left[ \frac{c(1 - \cos a)}{ \beta (1 + \alpha)} - \frac{\alpha \sin a}{\beta (1 + \alpha)} \right] \cos \xi + \left[ \frac{c \sin a}{ \beta (1 + \alpha)} + \frac{\alpha (1 - \cos a)}{ \beta (1 + \alpha)} \right] \sin \xi.  \label{Vpulsol}
\end{align}
Thus, we can fully characterize the traveling pulse solutions to (\ref{ringad}) as they depend upon model parameters. In the specific case of a cosine weight (\ref{cos}) and Heaviside firing rate, we note that the speed does not depend at all upon the threshold $\theta$ and the width does not depend upon the adaptation strength $\beta$ as shown in (\ref{adpulspeed}) and (\ref{adpulwid}). We demonstrate these relationships in Fig. \ref{expulplot}.

\section{Stability of the traveling pulse} \label{stabpulse}
We now calculate the stability of the traveling pulses derived in the previous section. To do so, we examine the evolution of small, smooth perturbations to the traveling pulse solution, using $(u,v) = (U (\xi), V (\xi)) +  \ve ( \bar{\psi}( \xi , t) , \bar{\phi}( \xi, t))$. Plugging this expansion into (\ref{ringad}) and truncating at ${\mc O}( \ve )$, we obtain
\begin{subequations}  \label{stablin}
\begin{align}
\frac{\pd \bar{\psi} (\xi,t)}{\pd t}-c\frac{\pd \bar{\psi} (\xi,t)}{\pd \xi} &= - \bar{\psi} (\xi,t) -\beta \bar{\phi} (\xi,t) + \int_{- \pi}^{\pi} w(\xi - y) f'(U(y)) \bar{\psi} (y,t) \d y, \\
\frac{\pd \bar{\phi} (\xi,t)}{\pd t} -c\frac{\pd \bar{\phi} (\xi,t)}{\pd \xi} &=   \alpha \bar{\psi} (\xi,t) - \alpha \bar{\phi} (\xi,t).
\end{align}
\end{subequations}
Due to the linearity of (\ref{stablin}), we can use separation of variables to characterize solutions \cite{ermentrout98,folias04}. Doing so we should look for solutions of the form $(\bar{\psi}(\xi,t),\bar{\phi}(\xi,t))=(\psi(\xi),\phi(\xi))e^{\lambda t}$, so we have the eigenvalue problem
\begin{subequations} \label{lin}
\begin{align}
\lambda \psi(\xi)-c\psi'(\xi) &= - \psi(\xi) - \beta \phi(\xi) + \int_{- \pi}^{\pi} w(\xi - y) f'(U(y))\psi(y) \d y, \\
\lambda \phi(\xi)-c\phi'(\xi) &=   \alpha \psi(\xi) - \alpha \phi(\xi).
\end{align}
\end{subequations}
For an arbitrary firing rate function $f$, we can expand both spatial functions $(\psi,\phi)$ in the Fourier series (\ref{psiphiexp}), where the number of terms $N$ is determined by the expansion of $w(x)$. Coefficients in (\ref{psiphiexp}) are then computed by solving the linear system
\begin{subequations} \label{fcoefpstab}
\begin{align}
( \lambda + 1) {\mc A}_k - c k {\mc B}_k + \beta {\mc M}_k &= w_k \int_{- \pi}^{\pi} \cos (k \xi) f'(U(\xi)) \psi ( \xi ) \d \xi, \\
( \lambda + 1) {\mc B}_k + c k {\mc A}_k + \beta {\mc N}_k &= w_k \int_{- \pi}^{\pi} \sin (k \xi) f'(U(\xi)) \psi ( \xi ) \d \xi, \\
( \lambda + \alpha) {\mc M}_k -c k {\mc N}_k &= \alpha {\mc A}_k, \\
( \lambda + \alpha) {\mc N}_k + ck {\mc M}_k &= \alpha {\mc B}_k. 
\end{align}
\end{subequations}
Solutions of the system (\ref{fcoefpstab}), along with their associated $\lambda$, are the eigensolutions to (\ref{lin}). To demonstrate these calculations, we consider the case of the cosine weight function (\ref{cos}). Here, we only have four coefficients ${\mc A}_1$, ${\mc B}_1$, ${\mc M}_k$, and ${\mc N}_k$ with four associated equations. Thus, by dropping the subscripts and applying self-consistency to (\ref{psiphiexp}) and (\ref{fcoefpstab}), we have the eigenvalue problem $\lambda \Phi = M \Phi$ where
\begin{align}
\Phi =  \left( \begin{array}{c} {\mc A} \\ {\mc B} \\ {\mc M} \\ {\mc N} \end{array} \right), \ \ \  M =  \left( \begin{array}{cccc} -1 + {\mc I}(\cos^2 \xi) & c + {\mc I} ( \cos \xi \sin \xi ) & - \beta & 0 \\ -c + {\mc I}( \cos \xi \sin \xi ) & -1 + {\mc  I} ( \sin^2 \xi ) & 0 & - \beta \\ \alpha & 0 & - \alpha & c \\ 0 & \alpha & -c & - \alpha \end{array} \right)  \label{spuleval}
\end{align}
where
\begin{align*}
{\mc I}( r(\xi )) = \int_{- \pi}^{\pi} r( \xi ) f'( U( \xi )) \d \xi,
\end{align*}
and the eigenvalue equation has a non-trivial solution when $\det (\lambda I - M) = 0$. To demonstrate this analysis, we consider the Heaviside firing rate function (\ref{H}), so that
\begin{align*}
{\mc I}( \cos^2 \xi )  = \frac{(1+\alpha)(1+ \cos^2 a)}{1 - \cos a}, \ \ \ \ \ \ \ & {\mc I}(\cos \xi \sin \xi ) = - \frac{(1+ \alpha) \cos a \sin a}{1- \cos a},  \\
  {\mc I}(\sin^2 \xi ) = \frac{(1+ \alpha) \sin^2 a}{1- \cos a}, \ \ \ \ \ \ \ & 
\end{align*}
where we have used
\begin{align*}
\frac{\d}{\d U} H(U( \xi ) - \theta) = \frac{\delta( \xi - \pi)}{|U'(\pi)|} + \frac{\delta(\xi - \pi + a)}{|U'( \pi -a )|}; \ \ \ \ \ \ \ \ -U'( \pi ) = U'( \pi -a) = \frac{1- \cos a}{1+ \alpha}.
\end{align*}
Therefore, we need to compute the roots $\lambda$ of the determinant
\begin{align}
\left| \begin{array}{cccc} \lambda + 1 - \frac{\D (1+ \alpha) ( 1+ \cos^2 a)}{\D 1- \cos a} & -c + \frac{\D (1+ \alpha) \sin a \cos a}{\D 1- \cos a} & \beta & 0 \\ c + \frac{\D (1+ \alpha) \sin a \cos a}{\D 1- \cos a} & \lambda + 1 - \frac{\D (1+ \alpha) \sin^2 a}{\D 1 - \cos a} & 0 & \beta \\ - \alpha & 0 & \lambda + \alpha & -c \\ 0 & - \alpha & c & \lambda + \alpha \end{array} \right|.
\end{align}
Upon applying the formula for the nonzero wave speed (\ref{adpulspeed}), we find the characteristic equation is given by
\begin{align*}
0 & = \lambda ( \gamma_1 + \gamma_2 \lambda + \gamma_3 \lambda^2 + \lambda^3), \\
\gamma_1 &= -\frac{4 \alpha (1+ \alpha) ( \beta - \alpha) \cos a}{1- \cos a}, \\
\gamma_2 &= \frac{2 \alpha [ 2(\beta - \alpha) - \cos a ((\beta - \alpha) + (1+ \beta))]}{1- \cos a}, \\  \gamma_3 &= - \frac{2 (1 + \alpha) \cos a}{1- \cos a}.
\end{align*}
Note that for $\beta> \alpha$ and $a = a_s \in [ \pi/2 , \pi]$, as in (\ref{adpulwid}), we have $\cos a_s < 0$, so $\gamma_1, \gamma_2, \gamma_3 > 0$. Thus, the three roots of $\gamma_1 + \gamma_2 \lambda + \gamma_3 \lambda^2  + \lambda^3$ must all have negative real part. For $a=a_u \in [0, \pi/2]$, then $\cos a_s > 0$, so $\gamma_1, \gamma_2, \gamma_3 < 0$, and at least one root has positive real part. Thus, as shown in \cite{sandstede07}, we can conclude that $(U_s,V_s)$ (respectively $(U_u,V_u)$) is asymptotically stable (unstable). In addition, we note that there is a one-dimensional nullspace of the linearized operator given in (\ref{lin}), associated with the eigenvalue $\lambda = 0$, given by $(U',V')$. We also recover the fact that for $\beta = \alpha$, we have $\lambda=0$ with multiplicity two (since $\gamma_1 = 0$ in that case).

\section{Noise-induced wandering of traveling pulses} \label{wanderpulse}
We will now examine the effect additive noise has upon the propagation of traveling pulses. Our approach will follow along similar lines to recent studies of stationary patterns \cite{hutt08}, traveling fronts \cite{bressloff12}, and stationary bumps \cite{kilpatrick13} in stochastic neural fields. However, the analysis must be generalized to a second-order system here. We start by supposing that we can track the position of the traveling pulse using a single stochastic variable $\Delta (t)$, representing stochastic motion of the pulse's position in the traveling wave coordinate $\xi$. Thus, as in \cite{armero98,bressloff12,kilpatrick13} we presume that the fluctuating terms in (\ref{radlang}) generate two phenomena that occur on disparate time scales. Motion of the pulse about its uniformly translating position occurs on long time scales, and fluctuations in the form of the pulse's profile occur at short time scales \cite{armero98,bressloff12}. Thus, we can express the solution $(u(x,t),v(x,t))$ to the vector system (\ref{radlang}) as the sum of a fixed traveling pulse profile $(U(\xi),V(\xi))$ displaced by stochastic variable $\Delta (t)$ and higher order time-dependent fluctuations $\ve (\Phi, \Psi) + \cdots$ in the profile of the pulse, so
\begin{subequations}  \label{vecexp}
\begin{align}
u(x,t) &= U( \xi - \Delta (t)) + \ve \Phi ( \xi - \Delta (t), t) + \cdots \\
v(x,t) &= V( \xi - \Delta (t)) + \ve \Psi ( \xi - \Delta (t), t) + \cdots 
\end{align} 
\end{subequations}
To linear order, the stochastic variables $\Delta (t)$ will obey Brownian motion, which we will calculate. By substituting the expressions (\ref{vecexp}) into (\ref{radlang}) and taking averages, we find that the ${\mc O}(1)$ deterministic system (\ref{adpulsys}) for $U$ and $V$ is satisfied. Proceeding to ${\mc O}(\ve)$, we find that
\begin{align}
\d \bp (\xi,t)  - {\mc L} \bp (\xi, t) \d t = \ve^{-1} \left( \begin{array}{c}U'(\xi)  \d \Delta (t) \\ V'(\xi )\d \Delta (t) \end{array} \right) + \left( \begin{array}{c} 0 \\ \d W (\xi , t) \end{array} \right)   \label{phipvec}
\end{align}
where $\bp (\xi,t) = \left( \Phi (\xi,t), \Psi (\xi,t) \right)^T $ and ${\mc L}$ is the linear operator defined
\begin{align*}
{\mc L} \mathbf{u} (\xi) = \left( \begin{array}{c} c u'(\xi)- u( \xi) - \beta v( \xi ) + \int_{- \pi}^{\pi} w(x-y) f'(U(y)) u( y ) \d y \\ cv'( \xi ) + \alpha u( \xi ) - \alpha v( \xi ) \end{array} \right) 
\end{align*}
for any vector $\mathbf{u}( \xi ) = ( u(\xi), v( \xi ))^T$ for integrable functions. Note that we can see the nullspace of ${\mc L}$ includes the vector $(U' ( \xi ), V'( \xi ))^T$ by differentiating (\ref{adpulsys}). To ensure solvability of (\ref{phipvec}), we require that the right hand side is orthogonal to all elements of the null space of the adjoint operator ${\mc L}^*$, which is defined using the inner product
\begin{align*}
\int_{- \pi}^{\pi} \mathbf{g}^T ( \xi) {\mc L} \mathbf{u} (\xi) \d \xi = \int_{- \pi}^{\pi} \mathbf{u}^T(\xi) {\mc L}^* \mathbf{g}(\xi) \d \xi,
\end{align*}
for any integrable vector $\mathbf{g} (\xi) = (g(\xi), h(\xi))^T$. It then follows
\begin{align}
{\mc L}^* \mathbf{g} = \left( \begin{array}{c} -c g'(\xi) - g(\xi) + \alpha h(\xi) + f'(U(\xi)) \int_{- \pi}^{\pi} w(\xi - y) g(y) \d y \\ -c h'(\xi) - \beta g(\xi) - \alpha h(\xi) \end{array} \right).  \label{puladjop}
\end{align}
We will solve for the null vector $( p (\xi), q (\xi))^T$ satisfying ${\mc L}^* (p,q)^T = (0,0)^T$ in a specific case in analysis that follows. Thus, we can ensure (\ref{phipvec}) has a solution by taking the inner product of both sides of the equation with $(p,q)^T$ to yield
\begin{align*}
\left[ \langle U' ( \xi ), p ( \xi) \rangle +\langle V' (\xi), q (\xi) \rangle \right] \d \Delta (t)  + \ve \langle \d W ( \xi, t), q ( \xi) \rangle = 0,
\end{align*}
which can be rearranged to yield
\begin{align}
\d \Delta = - \ve \frac{\langle \d W ( \xi, t), q ( \xi) \rangle}{ \langle U' ( \xi ), p ( \xi) \rangle +\langle V' (\xi), q (\xi) \rangle },  \label{delsde}
\end{align}
so $\d \Delta (t)$ is the increment of an ${\mc O}(\ve)$ white noise process with $\langle \Delta (t) \rangle = 0$ and $\langle \Delta (t)^2 \rangle  = \mathbf{D} t$ with
\begin{align*}
\mathbf{D} = \ve^2 \frac{\D \int_{- \pi}^{\pi} \int_{- \pi}^{\pi} q (x) q (y) \mathbf{C}(|x-y|) \d x \d y}{\left[ \D  \int_{- \pi}^{\pi} U'( x) p (x) \d x + \int_{- \pi}^{\pi} V'(x ) q (x) \d x \right]^2}.
\end{align*}
To calculate the effective variance $\langle \Delta (t)^2 \rangle$ of (\ref{delsde}), we now must determine the nullspace $(p,q)^T$ of the adjoint operator $\Lo^*$ given by (\ref{puladjop}), so
\begin{align*}
- c p'( \xi ) - p (\xi) + \alpha q (\xi) + f'(U(\xi)) \int_{- \pi}^{\pi} w( \xi - y) p (y) \d y &= 0 \\
-c q'( \xi ) - \beta p (\xi) - \alpha q (\xi) &= 0.
\end{align*}
For a Heaviside firing rate function (\ref{H}), cosine weight kernel (\ref{cos}), we have that the null vector $(p,q)^T$ must satisfy
\begin{subequations} \label{pulHadj}
\begin{align}
- c p'(\xi) - p( \xi ) + \alpha q( \xi ) + \frac{\delta ( \xi - \pi )}{|U'( \pi )|} \int_{- \pi}^{\pi} \cos ( \xi - y ) p(y) \d y \nonumber \\ 
+ \frac{\delta ( \xi - \pi + a)}{|U'( \pi - a)|} \int_{- \pi}^{\pi} \cos ( \xi - y ) p (y) \d y &= 0 \\
- c q'(\xi) - \beta p( \xi ) - \alpha q( \xi ) &= 0.  
\end{align}
\end{subequations}
A similar system was studied in \cite{kilpatrick08,ermentrout10} for an excitatory neural network on an infinite domain. Here, we must account for the periodic boundary conditions of the domain that cause exponentially decaying functions to wrap around. To proceed, note that for $\xi \neq \pi - a, \pi $, the system (\ref{pulHadj}) has solutions of the form $(p ( \xi ) , q( \xi ))^T = \vv \e^{\lambda \xi}$ with associated eigenvalue problem ${\mathbf A}^{\mathbf T} \vv = c \lambda \vv$ and $\mathbf A$ is defined in \eqref{MatrixLin}.
The associated eigenvalues are
\begin{align*}
\lambda = \lambda_{\pm} = \frac{m_{\pm}}{c} = \frac{1}{2c} \left[ -(1+ \alpha) \pm \sqrt{(1- \alpha)^2 - 4 \alpha \beta} \right],
\end{align*}
and the associated eigenvectors are
\begin{align*}
\vv_{\pm} = \left( \begin{array}{c} m_{\pm} + \alpha \\ - \beta \end{array} \right).
\end{align*}
Thus, due to the delta functions at $\xi = \pi -a, - \pi$, we expect the null-solution to (\ref{pulHadj}) to be of the form
\begin{align}
(p(\xi), q( \xi ))^T =& \gamma_+ \vv_+ \left[ {\mc H}(\xi ,  \pi , \lambda_+) + \chi {\mc H}(\xi, \pi - a, \lambda_+) \right] \e^{\lambda_+ \xi} \nonumber \\ & + \gamma_- \vv_- \left[ {\mc H}(\xi,  \pi , \lambda_-) + \chi {\mc H}(\xi,  \pi - a, \lambda_-) \right] \e^{\lambda_- \xi},  \label{pulHgenul}
\end{align}
where
\begin{align*}
{\mc H} (\xi,  \zeta, \lambda_{\pm}) = \left[ H(\xi - \zeta ) + \frac{\coth (-\lambda_{\pm} \pi ) -1}{2} \right] \e^{-\lambda_{\pm} \zeta},
\end{align*}
and the $\coth (-\lambda_{\pm} \pi )$ terms have arisen from an identity for the infinite series
\begin{align*}
 \sum_{n=1}^{\infty} \e^{2n \pi \lambda_{\pm}} = \frac{\coth( - \lambda_{\pm} \pi ) - 1}{2},
\end{align*}
necessary to account for the $2 \pi$ periodicity of the domain. We must choose the coefficients $\gamma_{\pm}$ such that the delta functions that arise from differentiating $(p(\xi),q(\xi))$ only appear in the $p ( \xi )$, so
\begin{align*}
\gamma_+ \vv_+ + \gamma_- \vv_- = \left( \begin{array}{c} \Gamma \\ 0 \end{array} \right),
\end{align*}
and we can take $\gamma_{\pm} = \pm 1$ so that $\Gamma = m_+ - m_- $. Now, in order to determine $\chi$, we substitute (\ref{pulHgenul}) into the system (\ref{pulHadj}), which generates the system
\begin{align*}
c ( m_+ - m_- ) &= - \nu_+^0 \left[ {\mc K} ( \pi , \lambda_+ )  + \chi {\mc K} ( \pi - a, \lambda_+ ) \right] + \nu_-^0 \left[ {\mc K} ( \pi, \lambda_- ) + \chi {\mc K} ( \pi -a , \lambda_-) \right], \\
\chi c ( m_+ - m_- ) &= - \nu_+^a \left[ {\mc K} ( \pi , \lambda_+ ) -  {\mc N} ( \pi , \lambda_+ )\tan a  + \chi \left( {\mc K} ( \pi - a, \lambda_+ ) -  {\mc N} ( \pi -a , \lambda_+ \right)\tan a \right] \nonumber \\
& + \nu_-^a \left[ {\mc K} ( \pi , \lambda_- ) -  {\mc N} ( \pi , \lambda_- ) \tan a + \chi \left( {\mc K} ( \pi - a, \lambda_- ) - {\mc N} ( \pi -a , \lambda_- \right)\tan a \right] ,
\end{align*}
where $\nu_{\pm}^x = (m_{\pm}+\alpha) \cos x/|U'(\pi-x)|$, and we can evaluate the functions
\begin{align}
{\mc K} ( \zeta, \lambda_{\pm} ) &=  \int_{- \pi}^{\pi} \cos( \xi) {\mc H}( \xi, \zeta , \lambda_{\pm} ) \e^{\lambda_{\pm} \xi} \d \xi = \frac{c^2 \sin \zeta + cm_{\pm} \cos \zeta}{(1+\alpha)(m_{\pm} + \alpha)}, \label{Kint} \\
{\mc N} ( \zeta, \lambda_{\pm} ) &= \int_{- \pi}^{\pi} \sin( \xi) {\mc H}( \xi, \zeta , \lambda_{\pm} ) \e^{\lambda_{\pm} \xi} \d \xi =  \frac{cm_{\pm} \sin \zeta - c^2 \cos \zeta}{(1+ \alpha) ( m_{\pm} + \alpha )}.  \label{Nint}
\end{align}
After simplifying considerably, we obtain the pair of equations
\begin{subequations} \label{puladjchi}
\begin{align}
m_+ - m_- &= \frac{m_+ ( 1+ \chi \cos a) - m_- ( 1+ \chi \cos a)}{1- \cos a},   \\
\chi ( m_+ - m_- ) &= \frac{\cos a (m_+ - m_-) [1+ \chi \cos a + \chi \tan a \sin a] }{1- \cos a},
\end{align}
\end{subequations}
and we have made use of the fact that $|U'( \pi )| = |U'( \pi - a)| = (1- \cos a)/(1 + \alpha)$, so (\ref{puladjchi}) is satisfied if $\chi = -1$.

\begin{figure}
\begin{center} \includegraphics[width=13cm]{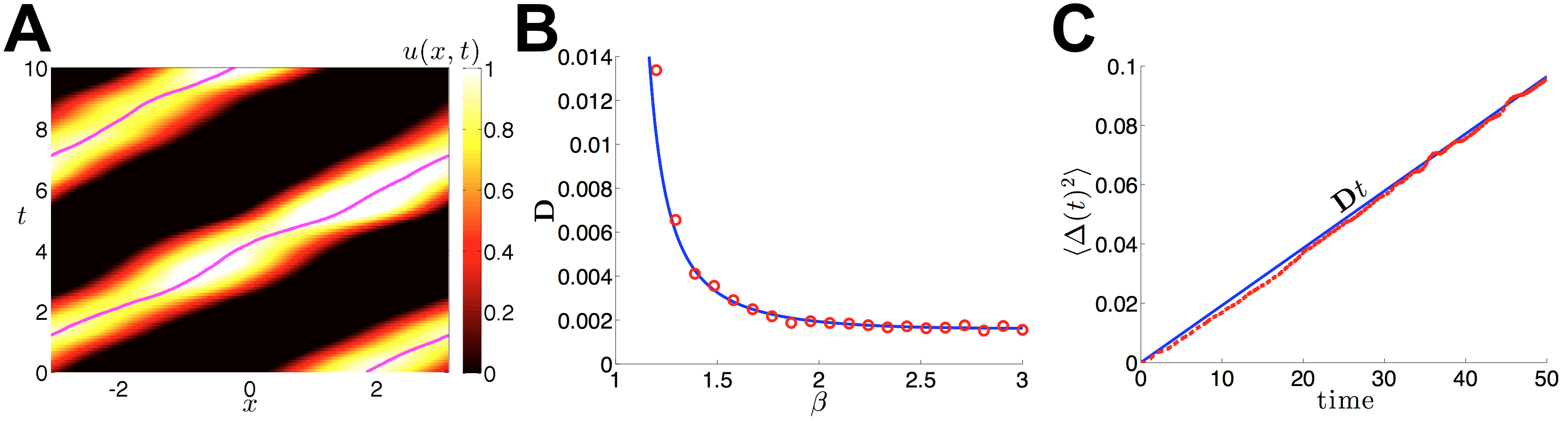} \end{center}
\caption{Wandering of traveling pulses in Langevin system (\ref{radlang}) due to noise with correlation $\mathbf{C}(x)= \cos (x)$. Firing rate function is Heaviside (\ref{H}), and weight is cosine (\ref{cos}). ({\bf A}) A single realization of neural activity $u(x,t)$, using the traveling pulse (\ref{Upulsol}-\ref{Vpulsol}) as initial condition. The superimposed line tracks the peak of the pulse. Parameters $\theta = 0.25$, $\alpha = 1$, and $\beta = 2$. ({\bf B}) The dependence of the diffusion coefficient on adaptation strength, computed using the asymptotic approximation (\ref{Hdiff}) (blue) and is computed numerically (red circles) across 1000 realizations run for 50 time units. ({\bf C}) Variance $\langle \Delta (t)^2 \rangle$ of the pulse's position $\Delta (t)$ computed across 1000 realizations (red dashed) scales linearly with time, as predicted by theory (blue solid). Diffusion coefficient is computed using (\ref{Hdiff}). Adaptation strength $\beta = 2$. Parameters $\ve = 0.03, \alpha = 1, \theta = 0.25$.}
\label{diffpulse}
\end{figure}

Now, with the adjoint vector (\ref{pulHgenul}) in hand, we can compute the diffusion coefficient describing the variance of pulse propagating in stochastic neural fields with Heaviside firing rate function (\ref{H}) and cosine weight (\ref{cos}). In the case of spatially structured additive noise with correlation function $\mathbf{C}(| x- y|) = \cos (x-y)$, we must compute
\begin{align}
\mathbf{D} &= \ve^2 \frac{\left[ \int_{- \pi}^{\pi} q (x) \cos x \d x \right]^2 + \left[ \int_{- \pi}^{\pi} q(x) \sin x \d x \right]^2}{\left[ \int_{-\pi}^{\pi} U'(x) p(x) \d x + \int_{- \pi}^{\pi} V'(x) q(x) \d x \right]^2}.  \label{DHcos}
\end{align}
To calculate (\ref{DHcos}), we will make use of the formulae (\ref{Kint}) and (\ref{Nint}). After some considerable calculations, we find that
\begin{align*}
\int_{- \pi}^{\pi} U'(x) p(x) \d x  &= - \frac{2 c ( m_+ - m_-) (1- \cos a)}{(1+ \alpha)^2}, \\
\int_{- \pi}^{\pi} V'(x) q(x) \d x &= - \frac{2c (c^2 - \alpha^2) (m_+ - m_-) (1- \cos a)}{(1+ \alpha )^2 (m_+ + \alpha)(m_- + \alpha)}, \\
\int_{- \pi}^{\pi} q(x) \cos (x) \d x &= \frac{\beta c(m_+-m_-)(\alpha(1-\cos a)-c\sin a)}{(1+ \alpha) (m_+ + \alpha)( m_- + \alpha)} , \\
\int_{- \pi}^{\pi} q(x) \sin (x) \d x &= \frac{\beta c(m_+-m_-)(c(1-\cos a)+\alpha \sin a)}{(1+ \alpha) (m_+ + \alpha)( m_- + \alpha)} , 
\end{align*}
so that we can compute
\begin{align}
\mathbf{D} =  \frac{\ve^2 \beta^3 (1+ \alpha)^2}{8 \alpha (1- \cos a) ( \beta - \alpha)^2}.  \label{Hdiff}
\end{align}
Thus, we have an asymptotic approximation for the ${\mc O}(\ve^2)$ effective diffusion coefficient $\mathbf{D}$ of a traveling pulse (\ref{Upulsol}-\ref{Vpulsol}). We demonstrate the accuracy of (\ref{Hdiff}) as compared to numerical simulations in Fig. \ref{diffpulse}. As predicted by our theory, averaging across numerical realizations of the Langevin equation (\ref{radlang}) shows the variance of the traveling pulse's position scales linearly in time.


\section{Discussion}
In this paper, we  have analyzed the effects of additive noise on traveling pulse solutions in spatially extended neural fields with linear adaptation.  We have considered random fluctuations by modeling the system as a set of spatially extended Langevin equations whose noise term is a $Q$-Wiener process that acts on the linear adaptation variable. Due to the translation invariance of the network, the noise-free system can support a continuum of spatially localized bump solutions. Bumps can be destabilized by increasing the strength of adaptation, leading to traveling pulse solutions. Near this criticality, we have derived a stochastic amplitude equation describing the dynamics of these pulses when the stochastic forcing and the bifurcation parameter are of comparable magnitude. Away from this bifurcation, we demonstrate numerically and analytically that noise causes traveling pulses to diffusively wander, so the variance of their position scales linearly with time.

We see several natural extensions of this work. First, it is not clear that noise should always be modeled as a $Q$-Wiener process. Noise could be temporally correlated or degenerate, acting only on some specific Fourier modes. For example, it would be interesting to derive amplitude equations in the simpler case of the ring model of orientations with degenerate noise, close to the deterministic pitchfork bifurcation that generates spatially localized bump solutions. Such degenerate noise may act only on the stable modes of the bump.
For stochastic PDEs, it has been shown both numerically \cite{roberts03} and theoretically \cite{blomker07} that degenerate noise can stabilize the dominant modes: noise can eliminate small linear instabilities.
Finally, it would be interesting to prove rigorous \emph{attractivity} and \emph{approximation} properties for the ansatz \eqref{ansatzSto} used in section \ref{stochcm}. That is, any solution to \eqref{slow} starting sufficiently close to the deterministic traveling pulse can be expanded in the form of the ansatz \eqref{ansatzSto} with $\Delta$ and $C$ satisfying the nonlinear stochastic differential equations \eqref{ampSto}, and that the residual parts of the ansatz remain small over a given fixed time window. Note that similar results have been obtained for the stochastic Swift-Hohenberg equations \cite{blomker01,blomker03,blomker05}.

As shown in this work, traveling pulses can be generated in one-dimensional adaptive neural fields. In addition, two-dimensional adaptive neural fields can support traveling spots \cite{coombes12b} and spiral waves \cite{huang04}. It may be possible to extend the analysis we have performed here to derive effective equations for the stochastic motion of spiral waves in the presence of noise. In this case, we would need to track how fluctuations alter the phase of the spiral wave as well as the position of the spiral's center. Furthermore, we could examine how such stochastic wave propagation interacts between multiple layers of a laminar neural field. Recently, such analysis has been carried out in the case that each layer supports a stationary bump, in the absence of noise \cite{kilpatrick13b}. Thus, it seems reasonable that, in the regime where coupling between layers scales with noise amplitude, a similar analysis could be extended to study how traveling pulses and spiral waves interact between layers.

\bibliography{wander}
\bibliographystyle{siam}

\end{document}